\documentclass[12pt,english]{article}
\usepackage{times}
\usepackage[T1]{fontenc}
\usepackage[latin1]{inputenc}
\usepackage{geometry}
\usepackage{amsmath}
\usepackage{color}
\usepackage{graphicx}
\usepackage{setspace}
\usepackage{amssymb}
 \usepackage{epstopdf}

\makeatletter

\providecommand{\LyX}{L\kern-.1667em\lower.25em\hbox{Y}\kern-.125emX\@}

 \newcommand{\lyxaddress}[1]{
   \par {\raggedright #1 
   \vspace{1.4em}
   \noindent\par}
 }

\usepackage{babel}
\makeatother
\begin{document}

\title{NOISE STABILIZED RANDOM ATTRACTOR}

\author{J. M. Finn{*}, E. R. Tracy{*}{*}, W. E. Cooke{*}{*}, and A. S. Richardson{*}{*}}

\maketitle

\lyxaddress{{*}T-15, Plasma Theory, Los Alamos National Laboratory; {*}{*}Department
of Physics, College of William and Mary}

\begin{abstract}
A two dimensional flow model is introduced with deterministic behavior
consisting of bursts which become successively larger, with longer
interburst time intervals between them. The system is symmetric in
one variable $x$ and there are bursts on either side of $x=0$, separated
by the presence of an invariant manifold at $x=0$. \textcolor{black}{In
the presence of arbitrarily small additive noise in the $x$ direction,
the successive bursts have bounded amplitudes and interburst intervals.
This system with noise is proposed as a model for edge localized modes
in tokamaks.} Further, the bursts can switch from positive to negative
$x$ and vice-versa. The probability distribution of burst heights
and interburst periods is studied, as is the dependence of the statistics
on the noise variance. The modification of this behavior as the symmetry
in $x$ is broken is studied, showing qualitatively similar behavior
if the symmetry breaking is small enough. Experimental observations
of a nonlinear circuit governed by the same equations are presented,
showing good agreement.
\end{abstract}

\section{Introduction}

This paper is \textcolor{black}{motivated by observations of extreme}
noise sensitivity in a two-dimensional flow of the form\begin{equation}
\frac{dx}{dt}=f(x,y)\equiv (y-1)x,\label{eq:x-eq.}\end{equation}
\begin{equation}
\frac{dy}{dt}=g(x,y)\equiv \epsilon y^{\nu }-x^{2}y.\label{eq:y-eq.}\end{equation}
This system is a low-dimensional model for the nonlinear behavior
of a plasma instability in which $y$ represents the pressure gradient,
and instability (with amplitude $x$) is driven by the pressure gradient
and fixed magnetic field line curvature. Such pressure -driven instabilities
are thought to be responsible for edge localized modes (ELMs) \textcolor{black}{observed
as fluctuations at the edge of a tokamak \cite{Zohm,Connor}. Some
ELMs, called Type-I ELMs, show temporal behavior which is quite simple,
consisting of well separated large bursts, indicating that their dynamics
can be represented by a low-order system. However, the time series
appear to show chaos, and it is of some interest to determine whether
this apparently chaotic behavior is indeed deterministic chaos or
whether it is due to sensitivity to noise from, for example, the plasma
core. For example, if the apparent chaos is due to noise, the behavior
can occur in a two dimensional model, whereas an autonomous model
showing similar apparently chaotic behavior must be at least three
dimensional.}

\textcolor{black}{The effect of noise has been studied in other experimental
physics situations, and the kind of extreme sensitivity to noise we
discuss here has been observed. For example, in experiments involving
the formation of droplets in a viscous fluid\cite{drop-falling-from-a-faucet},
the fluid is observed to form thin necks repeatedly as a part of the
process. Simulations showed the formation of necks, but the} \textcolor{black}{\emph{repeated}}
\textcolor{black}{formation of necks required noise in the modeling,
although extremely small noise gave agreement. Another example involves
studies of a Nd:YAG (neodymium doped yttrium aluminum garnet) laser
with an intercavity KTP (potassium titanyl phosphate) crystal. Theoretical
studies were performed to model the laser dynamics\cite{noise-on-chaotic-laser},
showing that the type-II chaotic dynamical behavior of the laser was
observed to be very sensitive to noise and was actually found to amplify
the noise. Because of the role of a very low level of noise in such
disparate physical systems, we have been motivated to do detailed
studies of (\ref{eq:x-eq.}), (\ref{eq:y-eq.}) and related systems
perturbed with a low level of noise.}

For the system \textcolor{black}{(\ref{eq:x-eq.}), (\ref{eq:y-eq.})}
with zero noise, $x$ grows if $y>1$, but for large enough $x$ the
term $-x^{2}y$, which represents the flattening of the pressure gradient
due to the fluctuation, enters. This causes a decrease in $y$, which
quenches the growth. For this flow, $x=0$ is an invariant manifold,
and is in fact the unstable manifold of a fixed point at $x=y=0$.
See Fig.~1. The $x-$axis is also an invariant
manifold, the stable manifold of the same fixed point. There are two
\textcolor{black}{unstable spirals} with $x=\pm x_{0}=\pm \sqrt{\epsilon }$.
The nonlinear deterministic behavior consists of spirals coming out
of the fixed points with $x=\pm x_{0}$, coming closer to the two
invariant axes on each pass, and developing \textcolor{black}{increasingly
larger bursts, one for each encircling of the spirals, more widely
separated in time.} Because of symmetry in $x$, identical bursts
can occur on both sides of $x=0$, isolated from each other by the
invariant manifold $x=0$.

With a small amount of uncorrelated Gaussian noise added to eq.~(\ref{eq:x-eq.}),
we find that %
\begin{figure}
\begin{center}\includegraphics[  width=3in,
  keepaspectratio]{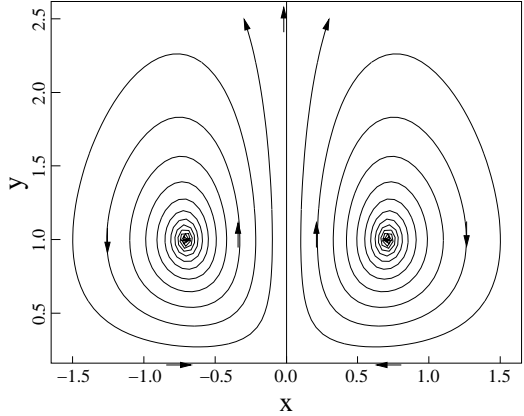}\end{center}

\caption{Orbits initiated near the fixed points at $x=\pm x_{0}=\pm \sqrt{\epsilon }$,
$y=1$. The orbit on the right spirals out clockwise, the one on the
left counter-clockwise. The $x-$and $y-$axes are, respectively,
stable and unstable manifolds of the fixed point at the origin.}
\end{figure}
the resulting nonlinear stochastic equation has the following property:
the bursts saturate in amplitude, leading to behavior that is qualitatively
similar to deterministic chaos. \textcolor{black}{We call this behavior}
\textcolor{black}{\emph{noise-stabilization}}\textcolor{black}{.}
Further, the noise allows transitions across the $y-$axis, an invariant
manifold for the deterministic system. Statistically, the dynamics
is sy\textcolor{black}{mmetric. In particular, we focus on the fraction
of the} number of bursts with $x<0$ compared with those with $x>0$;
with statistical symmetry these are equal. In the physical system
motivating this work, the processes we model as noise have a much
shorter correlation time than the processes described by the deterministic
equations (\ref{eq:x-eq.}), (\ref{eq:y-eq.}), \textcolor{black}{hence
modeling them as noise is appropriate. Noise-stabilized systems are
interesting for several reasons. Most importantly, although they can
exhibit dynamical behavior that is reminiscent of deterministic chaos,
it is likely that their behavior for very low noise level is distinguishable
from deterministic autonomous low dimensional systems. Our model system
was chosen to emphasize the noise-stabilizing effect, in the sense
that it has no attractor in the zero noise limit. In physical applications,
distinguishing noise-stabilized behavior from more familiar types
of dynamics could be critical for understanding and predicting how
the system under study will change as the noise driving is modified.}

There have been several related papers on nonlinear stochastic equations
which are sensitive to a small amount of noise. Sigeti and Horsthemke
\cite{sigeti} studied the effect of noise at a saddle-node bifurcation,
and found noise induced oscillations at a characteristic frequency.
Stone and Holmes \cite{Stone-and-Holmes} studied systems with an
attracting homoclinic orbit or an attracting heteroclinic cycle (structurally
stable because of the presence of a symmetry) in the presence of noise.
They found that the effect of the noise is to prevent the time between
bursts from increasing on each cycle. Stone and Armbruster \cite{Stone-Armbruster}
studied structurally stable (again because of symmetry) heteroclinic
cycles in the presence of noise, and analyzed the jumping between
invariant subspaces of the deterministic system. Armbruster and Stone
\cite{Armbruster-Stone} studied heteroclinic networks in the presence
of noise, and the induced switching between cycles. References \cite{Stone-and-Holmes,Stone-Armbruster,Armbruster-Stone}
stressed the importance of the linear part of the flow near the saddles.
Moehlis \cite{Moehlis} has investigated a system representing binary
fluid convection, and found that states with large bursts can be very
sensitive to noise. \textcolor{black}{References\cite{billings-schwartz,bollt-billings-schwartz}
deal with a system (SEIR or susceptible-exposed-infected-recovered)
describing epidemic outbreaks and show that chaos can be induced for
parameters far from the region for} which the deterministic system
is chaotic. 

The difference between our work and this previous work is the following.
Our work concerns a system which, in the absence of noise, has successive
bursts, each larger than its predecessor and separated by lengthening
time intervals. In the presence of noise, our system exhibits a finite
characteristic scale for the burst amplitude, a characteristic time
for bursts, and random switching across an invariant manifold of the
deterministic system. Further, our deterministic system is two-dimensional,
and therefore cannot have deterministic chaos, but the noise introduces
behavior which resembles deterministic chaos in several ways. In Refs.
\cite{Stone-and-Holmes,Stone-Armbruster,Armbruster-Stone} systems
with homoclinic or heteroclinic cycles were studied; the noise was
found to induce switching between subspaces and introduced a characteristic
time scale for intervals between bursts, but the bursts in the deterministic
system were limited in magnitude. The model of Ref.~\cite{Moehlis}
is four-dimensional, and therefore can, unlike our system, exhibit
chaotic behavior even without noise, in principle. It was found that
this specific system can have periodic bursts of infinite magnitude.
These infinite bursts are periodic in the sense that if the origin
and \textcolor{black}{infinity} are mapped to each other in a specific
way, the solutions to the equations can reach the origin in finite
time and can be integrated through it, leading to a periodic signal.
These states with large periodic bursts were found to be sensitive
to noise. This behavior is to be contrasted with the behavior we have
found from eqs.~(\ref{eq:x-eq.}), (\ref{eq:y-eq.}), in which (for
$\nu <2$) successive bursts get larger in magnitude, but no single
burst goes to infinity, and noise causes the bursts to behave in a
way that resembles deterministic chaos. The model in Refs.~\cite{billings-schwartz,bollt-billings-schwartz}
exhibits noise-induced chaos because of bi-instability, related to
the presence of two nearby unstable orbits.

The model we introduce is similar to the models of Refs. \cite{Stone-and-Holmes,Stone-Armbruster,Armbruster-Stone}
with a heteroclinic connection, in the formal sense that in our model
the $y-$axis is a heteroclinic orbit between the saddle at $(x,y)=(0,0)$
and the point at infinity. After a change of variables, the point
at infinity can be mapped to a finite point and the origin can be
left fixed. The new unstable manifold maps from the origin to this
second fixed point. However, additive noise in our system would then
map to non-additive noise in the compactified version. In particular
the noise disappears at the second fixed point, which is physically
unrealistic.

In Sec.~2 we introduce the deterministic form of the model and show
that with $\nu =1$ it is equivalent to the Lotka-Volterra predator-prey
model. We discuss the surface of section map $x\rightarrow x'=F(x)$,
taking minima of $x$ to maxima of $x$ (and vice-versa), as well
as the composite map $x\rightarrow x''$. 

In Sec.~3 we introduce the stochastic model and present results.
These results include those on the Lyapunov exponent $h_{1}$ and
the distribution of maxima of $|x|$ and the time interval $T$ between
bursts, and the dependence of these quantities on the noise diffusion
coefficient $D$. A brief discussion of the behavior near the $y-$axis
is shown. In this limit, the behavior in $x$ is linear and can be
treated by the Fokker-Planck equation, discussed in more detail in
Appendix A. 

In Sec.~4 we discuss the role of reflection symmetry in $x$ and
the effect of weak symmetry breaking. We also present results involving
modifications to the system at small and large $y$, and a modified
form of the equations in which the noise is replaced by a sinusoidal
perturbation. The results with an offset show that in a sense the
system with noise is structurally stable. The results with a sinusoidal
perturbation lend credence to the validity of the Lyapunov exponent
for the random case. 

In Sec.~5 we show results from an experiment with a nonlinear circuit,
showing noise stabilization in a physical system. 

In Sec.~6 we summarize our work.

\section{Deterministic model}

The deterministic form of the model we study is eqs.~(\ref{eq:x-eq.}),
(\ref{eq:y-eq.}). The parameters $\epsilon ,\: \nu $ are the only
parameters that cannot be removed by rescaling $x,$ $y,$ and $t$.
Starting with $x=0$ and $y>0$, $y$ increases in time, going to
infinity in finite time if $\nu >1$. \textcolor{black}{For $y>1$
small initial values of $x$ begin to grow. {[}The instantaneous growth
rate of $x$ in (\ref{eq:x-eq.}) equals $y-1$.{]}} If $x$ grows
at a rapid enough rate relative to $y$ (to be quantified later),
the second term in (\ref{eq:y-eq.}) eventually dominates the first
and $y$ decreases. For $\nu =1$ the system (\ref{eq:x-eq.}), (\ref{eq:y-eq.})
is the Lotka-Volterra predator-prey model. The usual form\cite{Strogatz}
of this system, in scaled variables, is\[
\frac{dX}{ds}=X(Y-1),\]
\[
\frac{dY}{ds}=(E-X)Y.\]
\textcolor{black}{With the change of variables} $X=x^{2}/2,\: Y=y,\: s=2t,\: E=\epsilon /2$,
it can be put in the form of eqs.~(\ref{eq:x-eq.}), (\ref{eq:y-eq.})
with $\nu =1$. Notice that in this latter form there is a symmetry
$x\rightarrow -x$ not present in the usual form. For this value of
$\nu $, equations (\ref{eq:x-eq.}) (\ref{eq:y-eq.}) can be written
in terms of $q=\ln x$, $p=\ln y$ in the form\[
\frac{dq}{dt}=e^{p}-1,\]
\[
\frac{dp}{dt}=\epsilon -e^{2q}.\]
\textcolor{black}{Thus eqs.~(\ref{eq:x-eq.}), (\ref{eq:y-eq.})
are an autonomous Hamiltonian system, with canonical variables} $(q,p)$:\begin{equation}
H(q,p)=e^{p}-p+\frac{1}{2}e^{2q}-\epsilon q=y-\ln y+\frac{1}{2}x^{2}-\epsilon \ln x.\label{eq:hamiltonian}\end{equation}
\begin{figure}
\begin{center}\includegraphics[  width=3in,
  keepaspectratio]{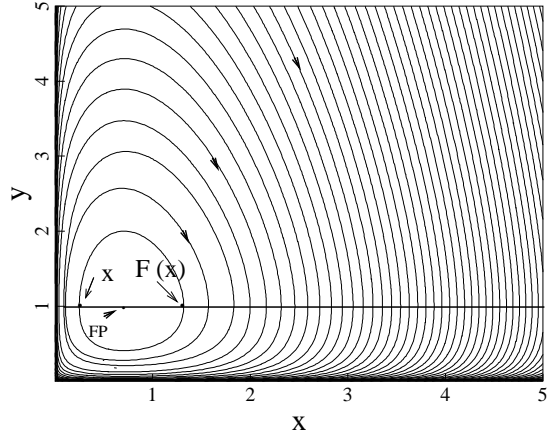}\end{center}

\caption{Contours of the Hamiltonian (\ref{eq:hamiltonian}) in $x,y$ for
the Lotka-Volterra model {[}$\nu =1$ in eq.~(\ref{eq:y-eq.}){]},
showing the fixed point at $(x,y)=(\sqrt{\epsilon },1)$ (labeled
FP) and the surface of section map $x\rightarrow F(x)$. For $\nu <1$
the orbits spiral into the fixed point; for $1<\nu <2$ the orbits
spiral out for all time; for $\nu >2$ the orbits spiral out, but
as soon as they cross $y$ with a small enough value of $x$ they
go off to infinity in one pass. See Sec.~2.2.}
\end{figure}
 Successive intersections of $H=const.$ with $y=1$ define a $1D$
surface of section \textcolor{black}{map $x\rightarrow x'=F(x)$}.
See Fig.~2. There are \textcolor{black}{centers at} $x=\pm x_{0}=\pm \sqrt{\epsilon },\: y=1$.
The mapping $F$ is determined \textcolor{black}{from} $H(q,p)$,
i.e.\begin{equation}
\frac{1}{2}x^{2}-\epsilon \ln x=\frac{1}{2}x'^{2}-\epsilon \ln x'.\label{eq:LV-map}\end{equation}
 For small $x$ we find $x\approx x'\exp (-x'^{2}/2\epsilon )$, which
can be approximated further by $x'=\sqrt{-2\epsilon \ln x}$. Thus
for small $x$ or very large $x'$, $F(x)$ is logarithmic in nature.
For large $x$ or small $x'$ we have the inverse $x'=x\exp (-x^{2}/2\epsilon )$. 

On the other hand, for $\nu >1$ the system is not Hamiltonian. It
has fixed points at $y=1$, $x=\pm x_{0}\textcolor {red}{=\pm \sqrt{\epsilon }}$
and at $x=y=0$. Near these fixed points, orbits evolve according
to the Jacobian $\mathsf{J}(x,y)=\nabla \mathbf{f}$, i.e.\begin{equation}
\frac{d}{dt}\delta \mathbf{x}(t)=\mathsf{J}\delta \mathbf{x}(t).\label{eq:variational}\end{equation}
For eqs.~(\ref{eq:x-eq.}), (\ref{eq:y-eq.}),\[
\mathsf{J}(x,y)=\left[\begin{array}{cc}
 y-1 & x\\
 -2xy & \epsilon \nu y^{\nu -1}-x^{2}\end{array}
\right].\]
 For the two fixed points at $x=\pm x_{0},\: y=1$ the eigenvalues
satisfy $\lambda ^{2}-\epsilon (\nu -1)\lambda +2\epsilon =0,$ and
are complex with positive real parts (unstable \textcolor{black}{spirals})
for \begin{equation}
0<\nu -1<\sqrt{8/\epsilon }.\label{eq:bound-on-epsilon-nu}\end{equation}
 \textcolor{black}{}Orbits continue to spiral out for $\nu >1$. This
is demonstrated by showing that the Hamiltonian for the case $\nu =1$
in eq. ~(\ref{eq:hamiltonian}) is a Lyapunov function for $\nu \neq 1$.
To show this, we note\[
\frac{dH}{dt}=\frac{dx}{dt}\frac{\partial H}{\partial x}+\frac{dy}{dt}\frac{\partial H}{\partial y}=\epsilon \left(y-1\right)\left(y^{\nu -1}-1\right).\]
Thus, for $\nu >1$, $dH/dt>0$ and the orbits spiral outward for
all time, since $H$ has a minimum at $x=x_{0},y=1$. For $\nu <1$,
$dH/dt<0$ and the orbits spiral in to the fixed point.

\textcolor{black}{The system has another fixed point, but with non-analytic
behavior in $y$ for noninteger $\nu $, at $x=0,\: y=0$.} The axes
$x=0,\: y=0$ are invariant manifolds; we consider only $y>0$, and
for the noise-free case orbits with $x(0)>0$ remain in that quadrant.
\textcolor{black}{In the range of $\epsilon $} and $\nu $ given
in eq.~(\ref{eq:bound-on-epsilon-nu}), orbits spiral away from the
fixed points at $(\pm x_{0},1)$ {[}Fig.~1{]}, approaching the $x-$
and $y-$axes, as shown in Fig.~3, %
\begin{figure}
\begin{center}\includegraphics[  width=4.5in,
  keepaspectratio]{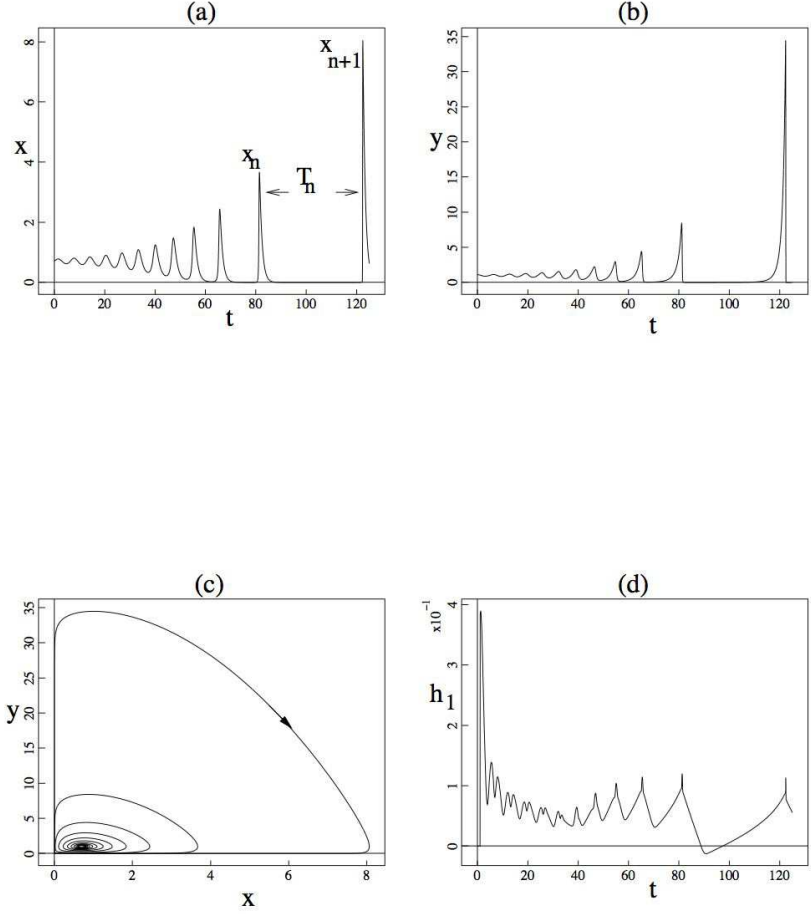}\end{center}

\caption{Orbits (a) $x(t),$ (b) $y(t)$ and (c) phase plane $y(x)$ for the
deterministic equations (\ref{eq:x-eq.}) and (\ref{eq:y-eq.}), with
$\epsilon =0.5$, $\nu =1.2$, with an initial condition near the
fixed point at $x=\sqrt{\epsilon },\: y=1$. The orbit spirals out
of the fixed point, continuing to expand, eventually piling up near
the invariant manifolds $x=0,\: y=0$, with bursts to large values
of $x$ and $y$ and long interburst time intervals spent mostly near
$x=y=0$. In (d) the finite time Lyapunov exponent $h_{1}(t)$ is
shown.}
\end{figure}
which has $\epsilon =0.5,\nu =1.2$. After an initial transient, the
motion is bursty, with each successive oscillation coming closer to
the axes, leading to a larger interburst interval, followed by a larger
burst. 

We compute the finite time Lyapunov exponent \begin{equation}
h_{1}(t)=\frac{1}{t}\ln \left(\frac{|\delta \mathbf{x}(t)|}{|\delta \mathbf{x}(0)|}\right),\label{eq:lyapunov}\end{equation}
where $\delta \mathbf{x}(t)$ is evolved according to eq.~(\ref{eq:variational})
and $\mathbf{x}(t)$ is evolved by eqs.~(\ref{eq:y-eq.}), (\ref{eq:x-eq.}).
In deterministic systems with a chaotic attractor, $h_{1}(t)$ measures
the average exponential rate of divergence, or stretching, over $0<t'<t$.
The largest Lyapunov exponent is the limit of $h_{1}(t)$ as $t\rightarrow \infty $
or the average, with suitable invariant measure, of $h_{1}(t)$ over
the attractor. \textcolor{black}{In this 2D system without time dependence
and with diverging orbits, the infinite time Lyapunov exponent does
not, strictly speaking, have significance. However, we will discuss
$h_{1}$ in more detail} in this section and Sec.~4.5, w\textcolor{black}{here
the orbits are bounded and it is therefore appropriate.} The exponent
$h_{1}(t)$ is shown as a function of time in Fig.~3d. It is clear
that $h_{1}(t)$ shows the bursts in $x$ and $y$, and decreases
whenever the orbit is near enough to the origin. In Fig.~4 we show
the zero contours of the larger eigenvalue $\rho (x,y)$ of the symmetrized
Jacobian $\mathsf{J}_{s}=(\mathsf{J}+\mathsf{J}^{T})/2$, computed
analytically.  %
\begin{figure}
\begin{center}\includegraphics[  width=4in,
  keepaspectratio]{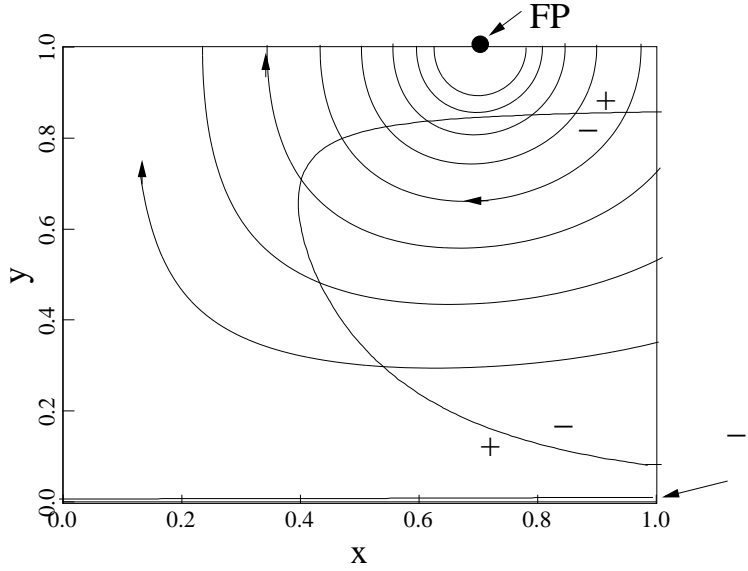}\end{center}

\caption{Zero contours of the larger eigenvalue $\rho $ of the symmetrized
Jacobian $J_{s}=(J+J^{T})/2$, showing the fixed point (FP) $(x,y)=(\sqrt{\epsilon },1)$,
the region ($+$) where $\rho >0$ and two regions ($-)$ where $\rho <0$,
one a very thin sliver near the $x-$ axis. Also shown is a representative
orbit spiraling out from the vicinity of the fixed point. The parameters
are as in Fig.~3. }
\end{figure}
This quantity is relevant because $|\delta \mathbf{x}(t)|=(\delta \mathbf{x}(t),\delta \mathbf{x}(t))^{1/2}$
evolves according to \[
(d/dt)(\delta \mathbf{x}(t),\delta \mathbf{x}(t))=(\delta \mathbf{x}(t),2\mathsf{J}_{s}\delta \mathbf{x}(t))\leq 2\rho (t),\]
so that $\rho (t)=\rho (x(t),y(t))$ is an upper bound for the local
contribution to $h_{1}(t)$, namely $(d/dt)\ln |\delta \mathbf{x}(t)|=|\delta \mathbf{x}(t)|^{-1}(d/dt)|\delta \mathbf{x}(t)|\leq \rho (t)$.
From this we find $dh_{1}(t)/dt\leq [\rho (t)-h_{1}(t)]/t$, $(d/dt)(th_{1})\leq \rho (t)$
or $h_{1}(t)\leq t^{-1}\int _{0}^{t}\rho (s)ds$. 

Further insight into the bursty nature can be obtained by finding
the surface of section, shown in Fig.~ 5 and discussed above for
the Hamiltonian case $\nu =1$. %
\begin{figure}
\begin{center}\includegraphics[  width=4in]{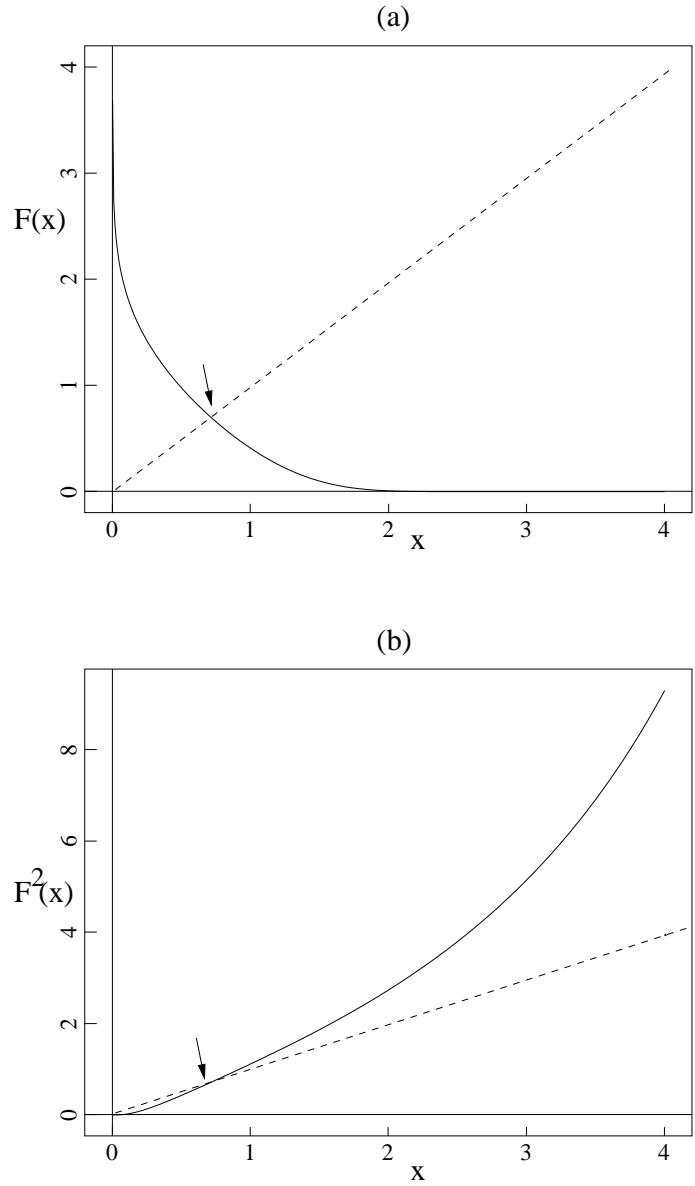}\end{center}

\caption{Surface of section (a) $x\rightarrow x'=F(x)$ from one crossing
of $y=1$ ($\dot{x}=0$) to the next, showing $x=\sqrt{\epsilon }$
as the fixed point. Parameters are as in Fig.~3. Composed surface
of section (b) $x\rightarrow x''=F^{2}(x)=F(F(x))$. The dashed lines
are, respectively, $x'=x$ and $x''=x$.}
\end{figure}
 For the parameters of Fig.~3, this map $x\rightarrow x'=F(x)$ is
shown in Fig.~5a. The slope $F'(x)$ at the fixed point $x=\sqrt{\epsilon }$,
computed numerically, equals $s_{1}=-1.17$. This value agrees with
the value obtained from the complex eigenvalues $\lambda $ of $\mathsf{J}(x_{0},y_{0})$,
which satisfy $\lambda =\lambda _{r}\pm i\lambda _{i}$ with $\lambda _{r}=\epsilon (\nu -1)/2$
and $\lambda _{i}=\pm \sqrt{2\epsilon }\left(1+O(\epsilon (\nu -1)^{2})\right)$,
which equals $\pm 1$ for $\epsilon =0.5$ and $\nu \ll 1$. This
gives $s_{1}\approx -e^{\epsilon (\nu -1)\pi /2}$. For $\epsilon =1/2,\: \nu =1.2$,
this gives $s_{1}=-1.17$, in agreement with the numerical results.
This value $s_{1}$ is less than $-1$, as it must be because the
fixed point is unstable. Note that the values of $x'$ for small $x$
rise rapidly as $x\rightarrow 0$ {[}$x'$ is approximately proportional
to $\sqrt{-\ln x}$, as suggested by the $\nu =1$ (Lotka-Volterra)
results discussed after eq.~(\ref{eq:LV-map}){]}, indicating that
orbits that are near $x=0$ when they pass $y=1$ lead to large succeeding
maxima. Even more pronounced is that for $x>3$ the values of $x'$
are vanishingly small, showing that moderately large maxima lead to
succeeding minima that are extremely close to the $y-$axis. In Fig.~5b
we show the composite surface of section $x\rightarrow x''$, from
one minimum to the next, or one maximum to the next. The slope at
the fixed point is $1.37\approx s_{1}^{2}$, as expected. For large
$x$, $x''=F^{2}(x)$ appears to be exponential in $x$.

Next, we turn to a discussion of the choice of the parameter $\nu $.
Let us investigate the range of the parameters $\nu ,\epsilon $ for
which the system exhibits successively larger, more widely separated
bursts.

Consider eqs.~(\ref{eq:x-eq.}), (\ref{eq:y-eq.}) for large $y$
and small $x$, i.e.\begin{equation}
\frac{dx}{dt}=yx,\label{eq:x-eq-y-large}\end{equation}
\begin{equation}
\frac{dy}{dt}=g(0,y)\approx \epsilon y^{\nu }.\label{eq:y-eq-y-large}\end{equation}
 From these we conclude \begin{equation}
x=x_{c}\exp \left[\frac{y^{2-\nu }}{\epsilon (2-\nu )}\right],\label{eq:large-y-solution}\end{equation}
where $x_{c}\exp \left[1/\epsilon (2-\nu )\right]$ is the value of
$x$ when the orbit passes $y=1$ with small $x$. Let us compare
the two terms on the right in eq.~(\ref{eq:y-eq.}), first for $\nu =1$
(Lotka-Volterra). The second term exceeds the first if $x^{2}>\epsilon $
and, since $x\sim e^{y/\epsilon }$, the nullcline $dy/dt=0$ is crossed,
and $y$ eventually decreases. For $1<\nu <2$, the nullcline is crossed
when $x^{2}\geq \epsilon y^{\nu -1}$ or\begin{equation}
x_{c}^{2}\exp \left[\frac{2y^{2-\nu }}{\epsilon (2-\nu )}\right]\geq \epsilon y^{\nu -1},\label{eq:nullcline}\end{equation}
which occurs eventually. So, in each burst, $y$ reaches a maximum
and begins to decrease, starting a new cycle, as long as $x\neq 0$.
(The orbits with $x=0$ go to infinity in finite time for $\nu >1$.)

For $\nu =2$, we can use eq.~(\ref{eq:x-eq-y-large}) with eq.~(\ref{eq:y-eq.})
for arbitrary $x$ (including the term $-x^{2}y$) to obtain, for
large $y$,\[
\frac{dy}{dx}=\epsilon \frac{y}{x}-x.\]
The solution is \[
y=\zeta x^{\epsilon }-\frac{x^{2}}{2-\epsilon },\]
with $\zeta >0$; the nullcline has $y=x^{2}/\epsilon $. For $\epsilon <2$,
the nullcline is crossed and the cycle begins again. For $\epsilon >2$
the nullcline is not crossed and the orbit can go off to infinity
in one cycle, in finite time.

For $\nu >2$, the nullcline in eq.~(\ref{eq:nullcline}) is never
reached if $x_{c}$ is small enough. This means that if the value
of $x$ when the orbit crosses $y=1$ is below some critical value,
the orbit will go off to infinity before another cycle. Therefore,
an orbit starting near the fixed point $(x,y)=(\sqrt{\epsilon },1)$
will encircle the fixed point a finite number of times and then go
off to infinity in finite time.

\section{Stochastic model and results}

\subsection{Model}

With noise, the system based on eqs.~(\ref{eq:x-eq.}), (\ref{eq:y-eq.})
is a nonlinear stochastic ODE, of the form\begin{equation}
\frac{dx}{dt}=f(x,y)+\sqrt{2D}\xi (t),\label{eq:x-with-noise}\end{equation}

\begin{equation}
\frac{dy}{dt}=g(x,y),\label{eq:y-with-noise}\end{equation}
with $\xi (t)$ representing uncorrelated unit variance Gaussian noise,
having $\langle \xi (t)\rangle =0,\: \langle \xi (t)\xi (t')\rangle =\delta (t-t')$.
Here, $D$ is the Brownian diffusion coefficient. For a low noise
level, $\xi (t)$ affects the dynamics only near the $y-$axis, where
$f(x,y)$ is small. The motivation for including noise in the $x-$equation
but not in the $y-$equation is the following. Without noise, when
the orbit is traveling along the $y-$axis for $y<1$, $x(t)$ can
decrease to a level that is unrealistically small for modeling any
physical application with noise. Noise prevents $x$ from becoming
so small for $0<y<1$, and therefore is expected to prevent the successive
bursts from continuing to increase in magnitude, with increasing interburst
time interval. We do not include noise in the $y-$equation because
noise could cause $y$ to become negative when the orbit is near the
$x-$axis. We will discuss a model allowing negative $y$ in Sec.~4.

We integrate the nonlinear stochastic ODE system (\ref{eq:x-with-noise}),
(\ref{eq:y-with-noise}) numerically, with a noise term in $x$ added
at each time step. Specifically, the time stepping from $t$ to $t+h$
is \begin{equation}
\begin{array}{c}
 x(t+h)=x(t)+hf\left(\frac{x(t)+x(t+h)}{2}\: ,\: \frac{y(t)+y(t+h)}{2}\right)+\sqrt{2Dh}\xi (t),\\
 y(t+h)=y(t)+hg\left(\frac{x(t)+x(t+h)}{2}\: ,\: \frac{y(t)+y(t+h)}{2}\right).\end{array}
\label{eq:finite-difference}\end{equation}
The implicit form of the deterministic part of the equations is solved
by a simple Picard iteration. The random term is added after this
iteration on the deterministic equations has converged. Each value
$\xi (t)$ is an independent random number with zero mean Gaussian
distribution and unit variance, and the coefficient $\sqrt{2Dh}$
is chosen \textcolor{black}{}to give results independent of the time
step $h$ (in a mean-square sense) in the limit $h\rightarrow 0$.

\subsection{Numerical results}

Results for the same parameters as in Fig.~3, with noise having $D=5\times 10^{-9}$,
are shown in Fig.~6, with $0\leq t\leq 1000$. %
\begin{figure}
\begin{center}\includegraphics[  width=3in,
  keepaspectratio]{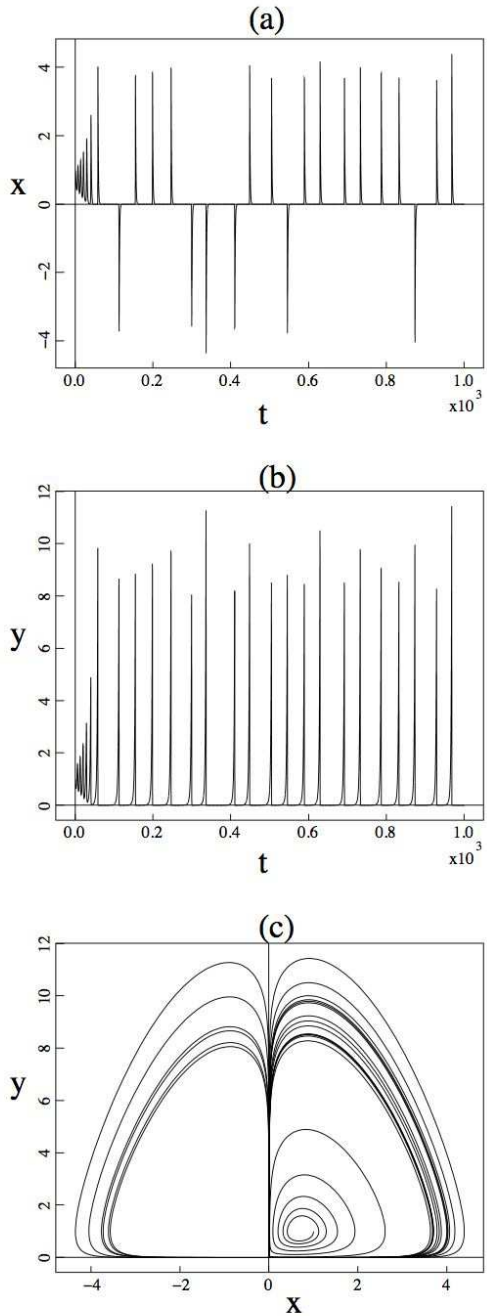}\end{center}

\caption{Orbits (a) $x(t),$ (b) $y(t)$ and (c) phase plane $y$ vs. $x$
for the system with noise, eqs.~(\ref{eq:x-with-noise}) and (\ref{eq:y-with-noise}).
The parameters are equal to those in Fig.~3, with $D=5\times 10^{-9}$.
The initial condition is near the spiraling fixed point, so that the
transient spiral shows. Note that the maximum time $t=10^{3}$ is
much larger than in Fig.~3. }
\end{figure}
The orbits are still of a bursty nature, but the bursts and the interburst
time intervals are limited in magnitude. The successive bursts appear
to be uncorrelated and bursts with $x$ negative are as common as
those with $x$ positive, after the transient near the fixed point
at $x=x_{0}=\sqrt{\epsilon },y=1$. To the eye, these results appear
similar to those of a chaotic deterministic system, e.g. the $y-z$
projection of the Lorenz system\cite{Ott}. %
\begin{figure}
\begin{center}\includegraphics[  width=3in,
  keepaspectratio]{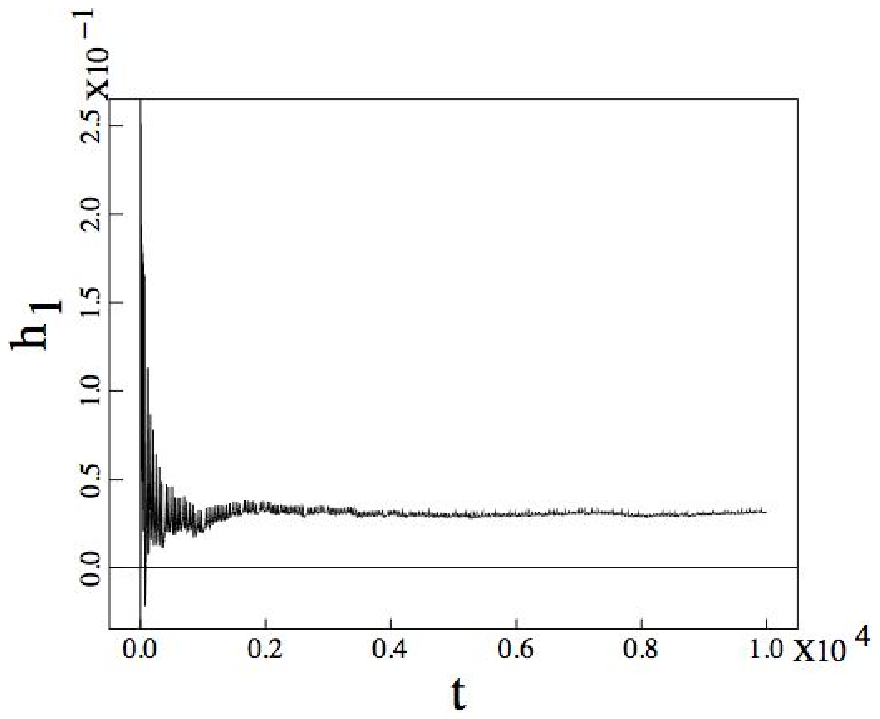}\end{center}

\caption{The finite time Lyapunov exponent up to $t=10^{4}$, for the case
of Fig.~6, showing a positive limiting value, $\lim _{t\rightarrow \infty }h_{1}(t)=0.032$.}
\end{figure}

In Figure 7 we show the finite time Lyapunov exponent $h_{1}(t)$
for the case of Fig.~6 for $0\leq t\leq 10^{4}$. The orbits $\mathbf{x}(t)=(x(t),y(t))$
given by eqs.~(\ref{eq:x-with-noise}), (\ref{eq:y-with-noise})
are affected by the noise $\xi (t)$ but the variational form for
$\delta \mathbf{x}(t)$ is eq.~(\ref{eq:variational}) and does not
directly involve the noise. {[}Two orbits $\mathbf{x}_{1}(t)$ and
$\mathbf{x}_{2}(t)=\mathbf{x}_{1}(t)+\delta \mathbf{x}(t)$ with slightly
different initial conditions are integrated in time with the same
realization of the noise $\xi (t)$.{]} For these parameters $h_{1}(t)$
converges to $0.032$ as $t\rightarrow \infty $. For several other
values of $\epsilon ,\nu ,$ and $D$, with $1<\nu <2$ and (\ref{eq:bound-on-epsilon-nu}),
similar results are obtained. This positive Lyapunov exponent shows
exponential divergence between nearby orbits. \textcolor{black}{This
suggests what appears to be evident from Fig.~6, namely that the
orbits behave chaotically. This conclusion is reasonable because the
system with noise is 2D with time dependence, and because the orbits
remain bounded for the time intervals studied, during which $h_{1}(t)$
appears to converge to a constant value. We will return to this discussion
in Sec.~4.5.}

To analyze the bursts in terms of amplitude and time interval between
bursts, we introduce $x_{n}$, $x_{n+1}$ and $T_{n}$. (See Fig.~3.)
These are, respectively, the amplitude (in $x$) of a burst (a local
maximum for positive $x$, a local minimum for negative $x$), the
amplitude of the following burst, and the time interval between them.
In Fig.~8 %
\begin{figure}
\begin{center}\includegraphics[  width=2.8in,
  keepaspectratio]{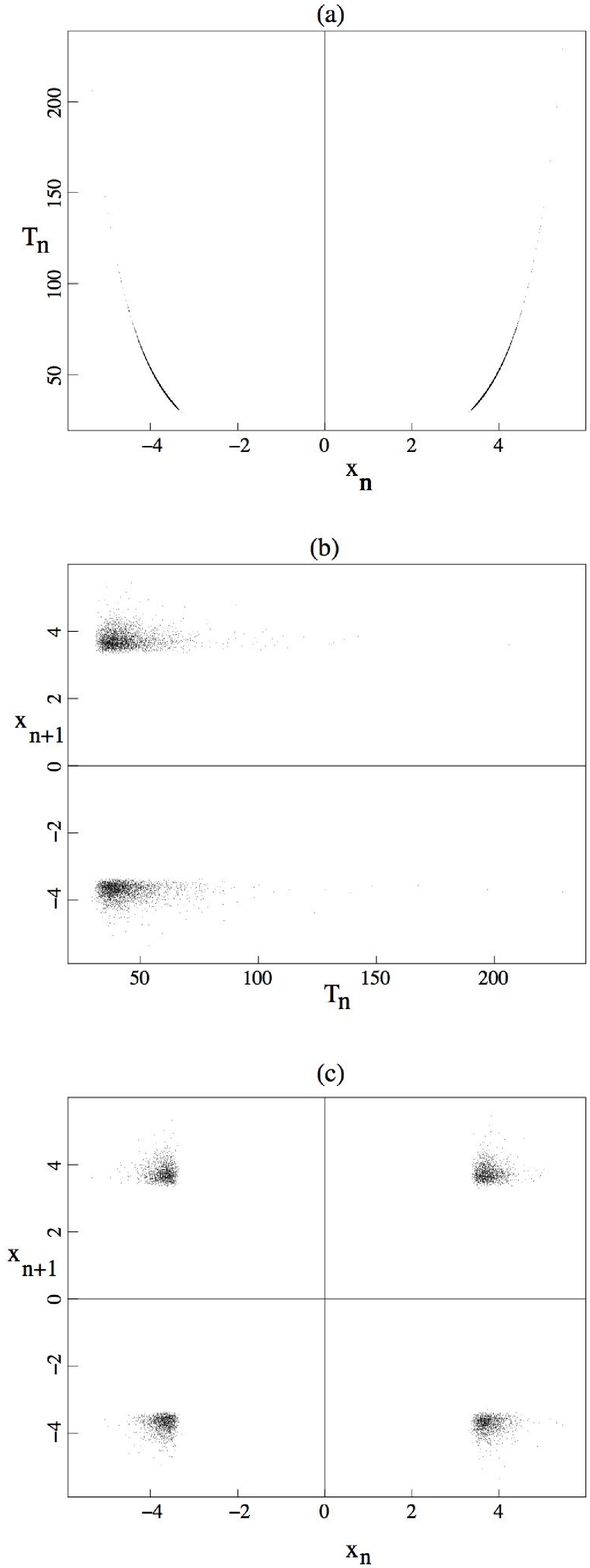}\end{center}

\caption{Scatter plots (a) $T_{n}$ vs. $x_{n}$, (b) $x_{n+1}$ vs. $T_{n}$
and (c) $x_{n+1}$ vs. $x_{n}$ for the case of Fig.~6. Note that
there is hardly any scatter in (a). The extent of the burst {[}measured
as $|x_{n}|$ or as the peak of $y(t)${]} determines $T_{n}$, because
after a larger burst the orbit approaches the origin closer to the
$x-$axis, because most of the interburst time is spent near $x=y=0$,
and because the noise is effective only near the $y-$axis. The statistics
plotted in (b) is symmetric in $x_{n+1}$ and has a long tail in $T_{n}$.
The plot in (c) is symmetric in $x_{n}$ and $x_{n+1}$, with four
essentially identical peaks near $|x_{n}|=|x_{n+1}|=4$. }
\end{figure}
we show scatter plots of $T_{n}$ vs. $x_{n}$, $x_{n+1}$ vs. $T_{n}$,
and the composite $x_{n+1}$ vs. $x_{n}$ for the parameters of the
case of Figs.~6 and 7, indicating the probability density functions
$f_{1}(x_{n},T_{n}),\: f_{2}(T_{n},x_{n+1})$ and $f_{3}(x_{n},x_{n+1})$.
These are the marginal distributions of the full distribution $g(x_{n},T_{n},x_{n+1})$
projected over $x_{n+1}$, $x_{n}$, and $T_{n}$, respectively. The
first has very little scatter. \textcolor{black}{This property is
related to two aspects. One is the fact that the noise is added only
to $x(t)$ and has little effect except when $x$ is small. The other
is that most of the time interval $T_{n}$ is spent near the saddle
at $x=y=0$, after the burst but before the orbit can be influenced
again by the noise, as it passes along the $y-$axis near $y=1$.}
This lack of scatter shows a very strong correlation. However, this
correlation is strongly nonlinear and would not be reflected in the
linear correlation coefficient, but would require a diagnostic such
as the conditional entropy \cite{Cover_Thomas}. The other plots show
the expected symmetry in $x$. Specifically, there are four equivalent
peaks in the four quadrants in Fig.~8c, showing that successive peaks
are positive or negative, independent of the sign of the previous
peak. Fig.~8b shows a long tail in $T_{n}$, and sharp cutoffs for
small $|x_{n}|$ and small $T_{n}$. 

In Fig.~9 %
\begin{figure}
\begin{center}\includegraphics[  width=3.5in,
  keepaspectratio]{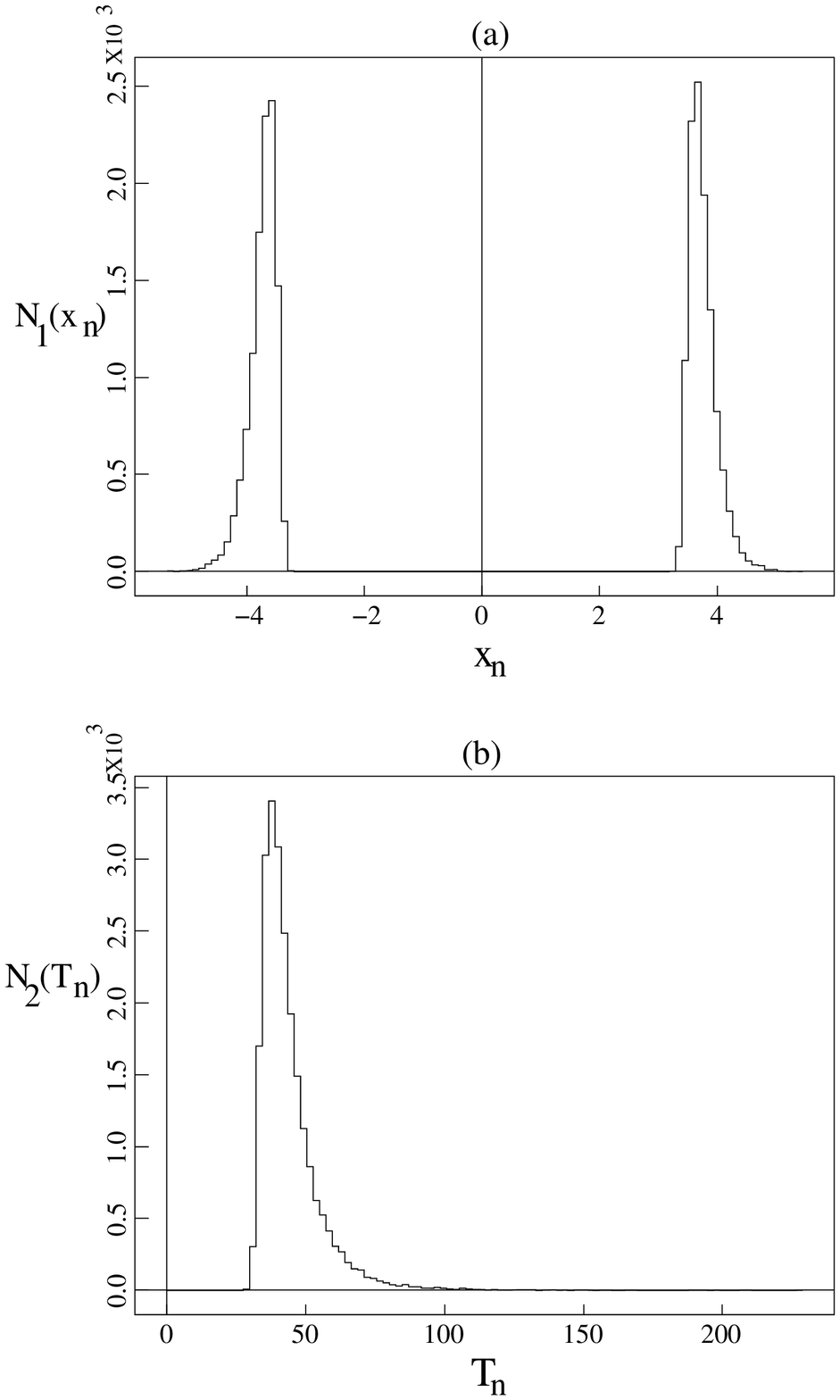}\end{center}

\caption{Histograms $N_{1}(x_{n}),N_{2}(T_{n})$ of (a) $x_{n}$, maxima of
$|x|$ and (b) time intervals $T_{n}$, respectively, showing the
marginal distributions for these quantities. The histogram of $|x_{n}|$
in (a) has tail with $|x_{n}|\gtrsim 5$ and a strong cutoff for $|x_{n}|<3.3$;
$T_{n}$ in (b) also has a tail to the right and a sharp cutoff to
the left. For this case the mean values are $\left\langle |x_{n}|\right\rangle =3.87$
and $\left\langle T_{n}\right\rangle =49.1$, respectively.}
\end{figure}
are histograms, showing the marginal distributions of $x_{n}$, at
the maxima of $\vert x\vert $, and the interburst time $T_{n}$.
(See Fig.~3.) The maximum time was $t=10^{6}$ and there were about
$23000$ peaks in $x_{n}$ and the same number of interburst intervals
$T_{n}$. The histogram of $x_{n}$ is symmetric and shows peaks at
$|x_{n}|=3.7$, with tails around $|x_{n}|=4.5$ and a sharp cutoff
inside at $|x_{n}|=3.3$. The latter histogram, reflecting the nonlinear
correlation of $T_{n}$ with $x_{n}$ shown in Fig.~8a, has a strong
cutoff inside $T_{n}=30$, a peak at $T_{n}=38$, and a tail for $T\sim 60-80$.

\textcolor{black}{Based on Sec.~2.2, we expect considerably different
results for $\nu >2$. These results show that, for the deterministic
system, if the value of $x$ at the} \textcolor{black}{\emph{throat}}
\textcolor{black}{$y=1$ is small enough, the orbit will go off to
infinity before another cycle occurs. Therefore, we expect that if
the noise level $D$ is small enough, the orbit may have a few bursts,
but will diverge to infinity as soon as the cycle comes close enough
to $x=0$ as it crosses $y=1$. For large values of $D$, the orbit
may behave as in Fig.~6 for a very long time, but whenever $x$ becomes
small enough at the inner crossing of $y=1$, the orbit will also
go to infinity before another cycle. Numerical simulations bear this
out.}

\subsection{Fokker-Planck analysis near $x=0$}

The peaks discussed in Figs.~8 and 9 are maxima in $|x|$, which
occur at $y=1$. These are related to the values of $x$ near zero
for which $y=1$: for small values of $D$, the noise is important
only near the $y-$axis, and as the orbit lifts off this manifold
it essentially obeys the deterministic equations, and therefore the
peaks in $|x|$ are determined to high accuracy by the crossing of
$y=1$ for small $x$. In this section we quantify this behavior by
means of analysis involving the Fokker-Planck equation for behavior
near the $y-$axis. 

As the orbit travels near the $y-$axis, $x(t)$ satisfies the linear
stochastic equation%
\begin{figure}
\begin{center}\includegraphics[  width=3in,
  keepaspectratio]{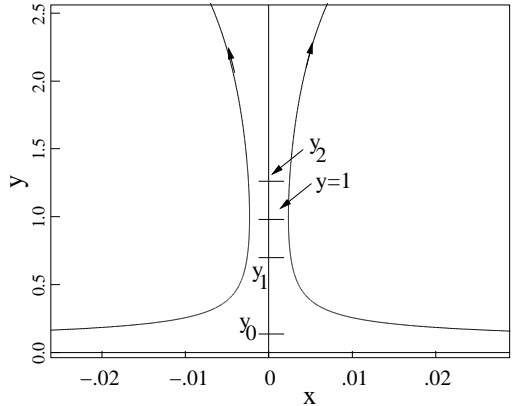}\end{center}

\caption{Sketch of deterministic orbits near $x=0$. The minima of $|x|$
are at the throat $y=1$. In this region, the equations can be linearized
with respect to $x$ and noise can have a large influence. The values
$y=y_{1},1,y_{2}$ correspond to $t=t_{1},0,t_{2}$ in the text.}
\end{figure}
\begin{equation}
\frac{dx}{dt}=\gamma (t)x+\xi (t),\label{eq:linear-stochastic}\end{equation}
where $\gamma (t)=y(t)-1$; for small $x$, $y$ satisfies $\dot{y}=\epsilon y^{\nu }$,
independent of $x$. The noise $\xi (t)$ has the statistical characteristics
described after eqs.~(\ref{eq:x-with-noise}), (\ref{eq:y-with-noise}).
Linearization in $x$ holds for small $D$, up to the time when the
term $-x^{2}y$ in eq.~(\ref{eq:y-eq.}) becomes important. For low
noise level (small $D$), the successive bursts are large in magnitude,
leading to small values of $x$ on the next pass. On each successive
pass near $y=1$, the correlation with the previous peak of $|x|$
is lost, according to the results shown in Fig.~8. This behavior
is due to the fact that for $g(0,y)=\epsilon y^{\nu }$ with $\nu >1$,
$x$ becomes small enough to become dominated by the noise while $y<1$.

In Appendix A we have included an analysis based on the Fokker-Planck
equation for orbits near $x=0$, where eq.~(\ref{eq:linear-stochastic})
is valid. Conclusions based on this Fokker-Planck analysis and direct
simulations are the following. The mean value $\left\langle |x_{n}|\right\rangle $
(c.f. Fig.~9a) decreases with $D$. The dependence of this quantity
is shown as a function of $D$ in Fig.~11a. %
\begin{figure}
\begin{center}\includegraphics[  width=3in,
  keepaspectratio]{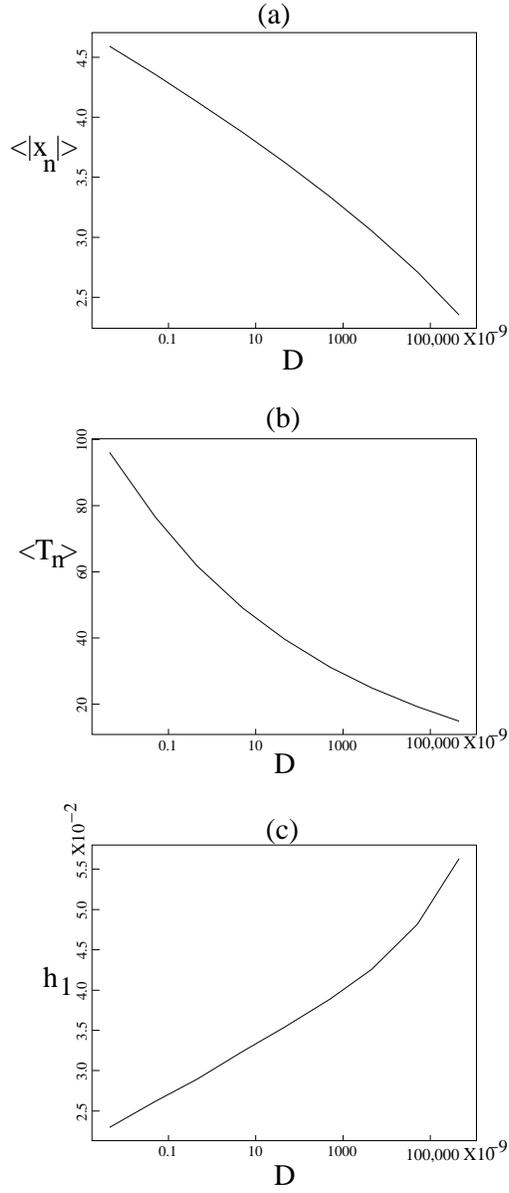}\end{center}

\caption{Mean values of (a) the burst peaks $\left\langle |x_{n}|\right\rangle $,
(b) the interburst time $\left\langle T_{n}\right\rangle $ and (c)
the Lyapunov exponent $h_{1}$ as functions of $D$. The parameters
(except for noise level) are the same as in Fig.~6. The quantities
$<|x_{n}|>$ and $h_{1}$ appear to behave logarithmically for small
$D$.}
\end{figure}
The mean of the histogram of the interburst time $T_{n}$ as a function
of $D$ is shown in Fig.~11b. The results for small $D$ in Fig.~11a
are qualitatively similar to the behavior of $F(x)$ shown in Fig.~5a.
This is expected because, as we have discussed in Appendix A, the
orbits cross $y=1$ with typical values of $x$ proportional to $\sigma _{x}\sim \alpha ^{1/2}\sim D^{1/2}/\epsilon ^{1/4}$,
and proceed with little subsequent effect of noise. The dependence
of $\left\langle |x_{n}|\right\rangle $ on $D$ appears to be approximately
logarithmic for small $D$, consistent with the approximately logarithmic
behavior of the map $F$ shown in Fig.~5a. It is also interesting
to note that, althoug\textcolor{black}{h $h_{1}$ increases with $D$,
the i}ncrease is logarithmic (for $D\lesssim 5\times 10^{-5}$) and
slow, varying by just over a factor of two for $5\times 10^{-12}<D<5\times 10^{-4}$.
This logarithmic behavior extrapolates to $h_{1}=0$ at the very low
level $D=10^{-19}$, giving $\sqrt{2Dh}=2\times 10^{-11}${[}c.f.
eq.~(\ref{eq:finite-difference}){]}. \textbf{\textcolor{black}{}}\textcolor{black}{Near
this value of $D$, $h_{1}$ appears to begin to diverge from logarithmic
behavior to remain positive. However, at these low noise values, roundoff
is comparable to the applied noise.}

The analysis in Appendix A shows that for small $x$, near the intersection
with $y=1$, $x$ has a Gaussian distribution, $f(x)\propto e^{-x^{2}/2\sigma _{x}^{2}}$
. This yields a distribution for $x'$, at the next crossing of $y=1$
where $|x_{n}|$ is a maximum, equal to\[
g(x')=\vert dx/dx'\vert f(x(x')),\]
where the functional form for $x(x')$ is shown in Fig.~5a. The second
factor is responsible for the sharp cutoff to the left o\textcolor{black}{f
the peak in $x'$ (Fig.~9a), corresponding to $x$ being in the tail
of the Gaussian. The} tail to the right of the peak in Fig.~9a is
due to the Jacobian factor $\vert dx/dx'\vert $. For example, for
$\nu =1.2$ the behavior for small $x$ from Fig.~5 is similar to
that for $\nu =1$, derived after eq.~(\ref{eq:LV-map}), namely
$x'\sim \sqrt{-\ln x}$. From the Gaussian form for $f(x)$ we obtain
$|dx/dx'|\sim x'e^{-x'^{2}}$ and\[
g(x')\propto \left(x'e^{-x'^{2}}\right)e^{-\frac{x(x')^{2}}{2\sigma _{x}^{2}}}.\]
The first (Jacobian) factor $x'e^{-x'^{2}}$ gives a Gaussian-like
tail for large $x'$ and the second factor gives a cutoff for $x'$
close to the fixed point $x'=x_{0}=\sqrt{\epsilon }$, where $x'-x_{0}=-s_{1}\left(x-x_{0}\right)$.
This cutoff is sharp if $\sigma _{x}\ll x_{0}$.

\section{\textcolor{black}{The role of symmetry and relation with other models}}

\textcolor{black}{We have commented that the system (\ref{eq:x-with-noise}),
(\ref{eq:y-with-noise}) has certain features that are not generic.
These issues are (a) the reflection symmetry of the equations in $x$;
(b) the fact that deterministic orbits eventually go to infinity,
and (c) the non-analytic behavior of $y^{\nu }$ near $y=0$. In this
section we discuss results obtained when the system is modified in
these areas. To deal with issue (a), we destroy the symmetry in $x$
by an offset {[}a constant term ad}ded to eq.~(\ref{eq:x-with-noise}){]}.
These results suggest a modification to the notion of structural stability
in the presence of noise: the behavior is qualitatively unchanged
if the offset is small relative to the noise. To deal with issue (b),
we show results in which the behavior for large $y$ is modified,
preventing orbits from going to l\textcolor{black}{arge $y$. Regarding
issue (c), we modify the system near $y=0$ to remove the non-analytic
behavior there.} \textcolor{red}{}\textcolor{black}{We also discuss
modifications breaking the reflection symmetry in $x$ together with
limiting the behavior for large $y$. Finally, we discuss modifications
to the system involving adding a sinusoidal perturbation to eq.~(\ref{eq:x-eq.})
in place of the noise term. In these studies, conventional deterministic
chaos, characterized by a Lyapunov exponent, is observed and compared
with results with noise. }

\subsection{Breaking of the symmetry in $x$}

We have investigated the effect of breaking the reflection symmetry
$x\rightarrow -x$ in eqs.~(\ref{eq:x-with-noise}), (\ref{eq:y-with-noise}),
motivated by the experimental results shown in Sec.~5.3. The simplest
way of breaking this symmetry is to introduce a constant offset. With
this offset, eq.~(\ref{eq:x-with-noise}) takes the form \begin{equation}
\frac{dx}{dt}=(y-1)x+a+\sqrt{2D}\xi (t),\label{eq:with-offset}\end{equation}
with the $y-$equation unchanged. \textcolor{black}{Numerical results
with zero noise show that for $a>0$ a stable limit cycle is formed
to the right of $x=0$, and points near $(x,y)=(0,0)$ go into this
limit cycle.} (For $a<0$ the results are identical, with $x\rightarrow -x$.)
Therefore the zero noise results of Sec.~2 are not structurally stable
with respect to such an offset.%
\begin{figure}
\begin{center}\includegraphics[  width=4.5in,
  keepaspectratio]{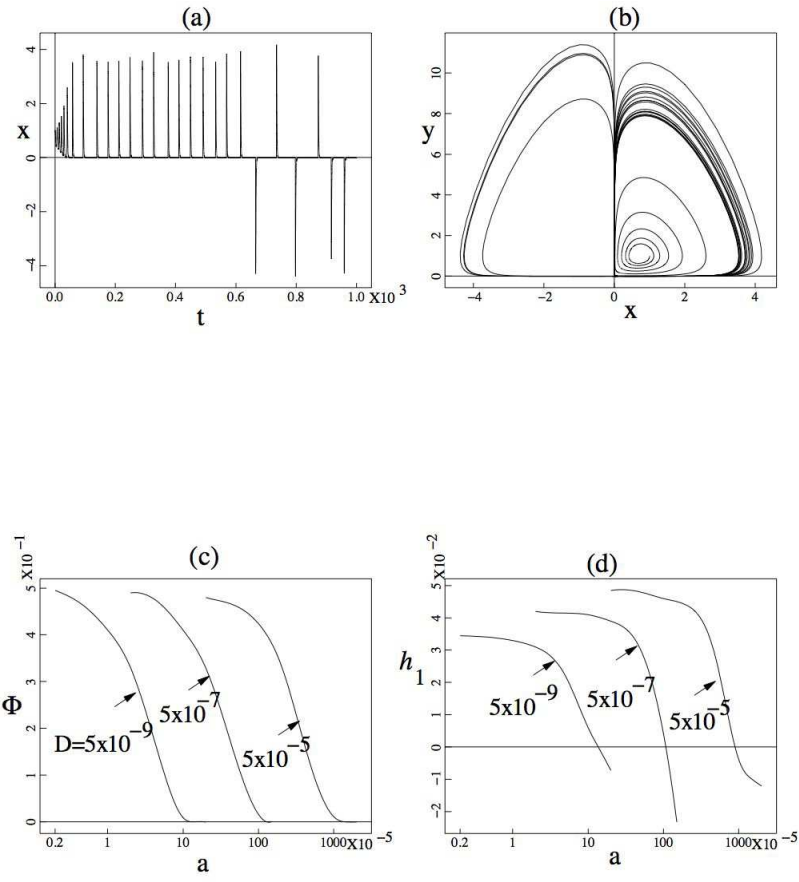}\end{center}

\caption{Results with an offset {[}c.f. $a$ in eq.~(\ref{eq:with-offset}){]}.
In (a), (b) are $x(t)$ and $y$ vs $x$ for parameters as in Fig.~6,
again with $D=5\times 10^{-9}$. In (c), (d) are the fraction $\Phi $
of bursts to the left and the Lyapunov exponent $h_{1}$, for three
values $D=5\times 10^{-9},5\times 10^{-7},5\times 10^{-5}$.}
\end{figure}

However, in the presence of noise, the results change considerably.
In Figs.~12a,b we show $x(t)$ and the phase portrait $y$ vs $x$
for a case with the same parameters as in Fig.~6 (in particular with
$D=5\times 10^{-9}$), but with $a=5\times 10^{-5}$. The results
are qualitatively similar to those in Fig.~6 except that most of
the bursts go to the right. In Fig.~12c we show the fraction $\Phi $
of bursts that go to the left as a function of the offset $a$ for
three values of $D$, and in Fig.~12d we show the Lyapunov exponent
$h_{1}$. For $a\lesssim \sqrt{D}$, $h_{1}$ and the fraction $\Phi $
are appreciable and the orbits behave qualitatively as in Fig.~6.
For $a\gtrsim \sqrt{D}$, on the other hand, \textcolor{black}{virtually
all the orbits go to the right} ($\Phi \approx 0$) and have negative
Lyapunov exponent and therefore behave qualitatively as the limit
cycle found for $D=0,\: a>0$. These results, and those of Appendix
A showing $\sigma _{x}\sim \sqrt{D}$, indicate that the offset changes
the results qualitatively if it moves the orbit outside the region
near $x=0$ where noise dominates.

This brings up the issue of structural stability of the behavior observed
for $a=0$. For zero noise, this behavior, seen in Fig.~3, is certainly
not structurally stable. However, for $D>0$ the qualitative behavior
persists as long as $a\lesssim \sqrt{D}$. In this modified sense,
the system with finite noise is structurally stable.

We will return to the issue of an offset in the electronic circuit
in the next section.

\subsection{Modifications for large $y$}

We have discussed the deterministic model for $\nu >2$ in Sec.~2,
showing that orbits go to infinity after a few passes near the fixed
point $(x,y)=(x_{0}=\sqrt{\epsilon },1)$. The dynamics in the presence
of noise is the following: if the noise is large enough, the value
of $x$ at the throat where $y=1$ will typically be large enough
that the system encircles $(x_{0},1)$ many times. Even with noise,
however, eventually an orbit comes through the throat with small enough
$x$ \textcolor{black}{for the system to go} to infinity before another
cycle can occur.

A system related to eqs.~(\ref{eq:x-eq.}), (\ref{eq:y-eq.}) with
orbits that do not to to infinity is the predator-prey system of Odell
\cite{Odell,Strogatz}. This system can be put into the form\[
\frac{dX}{ds}=X(Y-\eta ),\]
\[
\frac{dY}{ds}=Y^{2}(1-Y)-XY,\]
or by a change of variables ($X=\eta x^{2}/2,\: Y=\eta y,\: s=2t/\eta $)
\begin{equation}
\frac{dx}{dt}=(y-1)x,\label{eq:odell-1}\end{equation}
\begin{equation}
\frac{dy}{dt}=\epsilon y^{\nu }(1-\eta y)-x^{2}y,\label{eq:odell-2}\end{equation}
with $\epsilon =\nu =2$, i.e.~the form of eqs.~(\ref{eq:x-eq.}),
(\ref{eq:y-eq.}) with $\nu =2$ and $y^{2}\rightarrow y^{2}(1-\eta y)$.
This system has fixed points at $x=\pm \sqrt{\epsilon (1-\eta )},\: y=1$.
For $\epsilon =\nu =2$ these fixed points are unstable if $\eta <1/2$
and oscillating (complex eigenvalues) if $\eta <\sqrt{3}/2$. This
system also has a saddle at $x=y=0$, with zero eigenvalue in the
$y$ direction. In addition, it has a fourth fixed point, with $x=0$
and $y=1/\eta $. This fixed point is a saddle, stable in the $y-$direction
and unstable in the $x-$direction; the section of the $y-$axis with
$0<y<1/\eta $ is a heteroclinic line. Because of the presence of
this saddle, there are two stable limit cycles, related by the reflection
symmetry in $x$, to which typical orbits converge. For $\eta $ small,
this limit cycle has large excursions, with peaks in $y$ approaching
$1/\eta $. We have studied eqs.~(\ref{eq:odell-1}), (\ref{eq:odell-2})
with noise in $x$, and with $\epsilon $, $\nu $ in the range of
parameters of Fig.~3. The results are similar to those of (\ref{eq:x-with-noise}),
(\ref{eq:y-with-noise}), as long as $D$ is large enough that the
excursions almost always have $y\ll 1/\eta $. Specifically, the value
of $h_{1}$ and the probability density plots as in Figs.~8, 9 are
essentially identical. The effect of positive $\eta $ is similar
to the effect of \textcolor{black}{clipping the} voltage corresponding
to $y$ in the circuit (see Appendix B), except that by design the
clipping turns on much more rapidly than the factor $(1-\eta y)$
in eq.~(\ref{eq:odell-2}).

\subsection{Modifications near $y=0$}

We have studied the system (\ref{eq:x-with-noise}), (\ref{eq:y-with-noise})
with $\epsilon y^{\nu }\rightarrow g_{0}(y)=\epsilon (\beta y+y^{\nu })$.
This modification regularizes the vicinity of $y=0$: the saddle at
the origin is no longer dominated by $y^{\nu }$, and has eigenvalues
$-1,\: \epsilon \beta $. The spiraling fixed points have $x=\pm x_{0}=\pm \sqrt{\epsilon (1+\beta )},\: y=1$.
We have found that noise has the same qualitative influence for positive
$\beta $ as it does for $\beta =0$. In Fig.~13a %
\begin{figure}
\begin{center}\includegraphics[  width=4in,
  keepaspectratio]{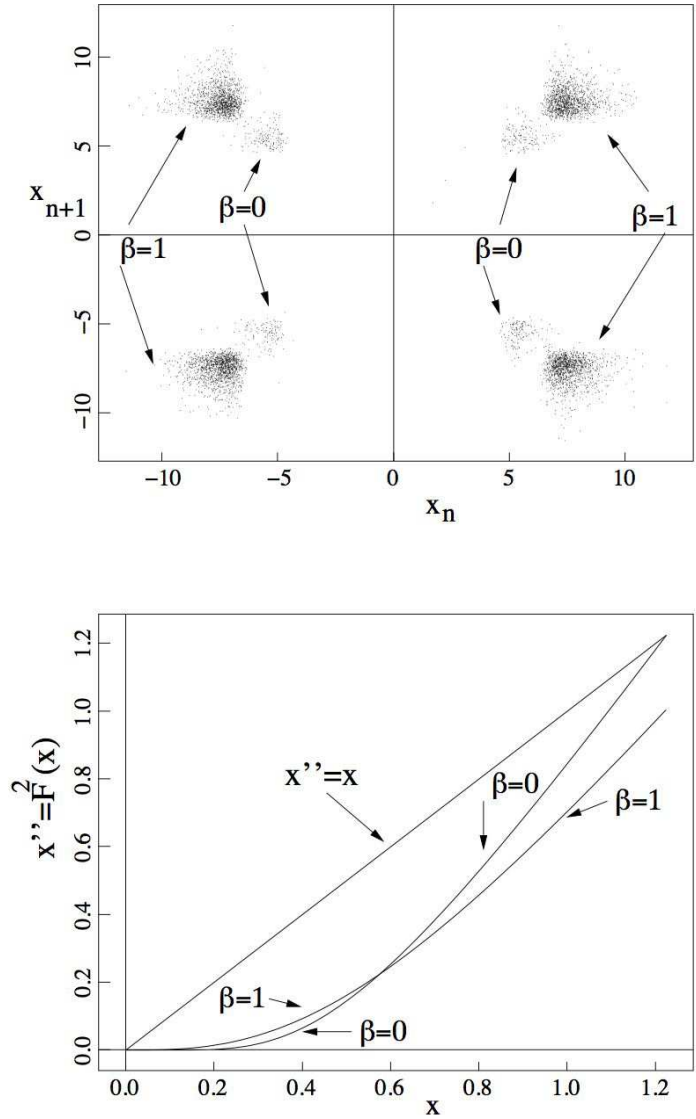}\end{center}

\caption{Scatter plot (a) $x_{n}\rightarrow x_{n+1}$ in the notation of Fig.~3
(successive maxima of $\vert x\vert $), for $\epsilon =1.5,\: \nu =1.2,\: D=5\times 10^{-5}$
and both $\beta =0$ and $\beta =1$, showing four-fold symmetry in
both cases. Surface of section $x\rightarrow x''=F^{2}(x_{n})$ (b)
for $0<x<\sqrt{\epsilon (1+\beta )}=1.22$ for the deterministic case
with $\epsilon =1.5$ and both values of $\beta $. For $\beta =1$
the fixed point is at $x=\sqrt{\epsilon (1+\beta )}=1.73$.}
\end{figure}
we show the sc\textcolor{black}{atter plot $x_{n}\rightarrow x_{n+1}$
for $D=5\times 10^{-5},\: \epsilon =1.5$, $\nu =1.2$, and $\beta =0$,
$\beta =1$ superimposed. Fo}r $\beta =1$ the eigenvalue $\epsilon \beta >1$,
which implies that, when following a deterministic orbit along the
$x-$ axis and up along the $y-$axis, it ends up further from the
$y-$axis than it started from the $x-$ axis. (For the equations
linearized about the origin, $x^{\epsilon \beta }y$ is constant.)
This is related to the liftoff phenomenon of Refs.\cite{Stone-Armbruster,Armbruster-Stone}.
Based on this consideration, one might expect that the sign of $x_{n+1}$
might correlate with the sign of $x_{n}$, and the symmetry of the
scatter plot would be broken, with the distribution $f_{3}(x_{n},x_{n+1})$
having more points in the NE and SW quadrants and fewer in the SE
and NW quadrants, while of course still preserving the symmetry in
the marginal distribution of $x_{n}$, $\int f_{3}(x_{n},x_{n+1})dx_{n+1}$.
Nevertheless, the scatter plot for $\beta =1$ appears to have the
same symmetry as for $\beta =0$.

This four-fold symmetry is explained by Fig.~13b, which shows the
surface of section $x\rightarrow x''=F^{2}(x)$ for $0<x<x_{0}$,
similar to that in Fig.~5b. For both cases $x''\ll x$. For these
parameters $x''\sim x^{3}$ for $\beta =1$, while $x''$ goes to
zero faster than any power when $\beta =0$. The origin is so very
attracting for $F^{2}$ because small $x$ maps to large $x'$ under
$F$ and the orbit from there passes extremely close to $y=0$, thereby
leading to extremely small $x''$ in spite of $\epsilon \beta >1$.
Because of this property, if an orbit starts with $x\sim \sigma _{x}$
at $y=1$ and executes one cycle, the value $x=x''$ when it crosses
$y=1$ after this cycle will be so small ($x''\sim \sigma _{x}^{3}$
for $\beta =1$) that it is dominated by the noise added for small
$x$ and will even for $\beta =1$ be nearly independent of $x$.
This four-fold symmetry was observed for these parameters for $5\times 10^{-13}<D<5\times 10^{-3}$.

Although the increase of $\beta $ has no effect on the symmetry of
the scatter plot $x_{n}\rightarrow x_{n+1}$, it has a profound influence
on the burst intervals $T_{n}$. For larger $\beta $, typical values
of $T_{n}$ (not shown) are much smaller because of the liftoff phenomenon.
This dependence of $T$ on $\beta $ is understood easily. Suppose
the orbit enters the region $[0,a]\times [0,b]$ with $y=y_{0}$.
We find that if $\beta y\gg y^{\nu }$, the time to exit the region
equals $T_{1}\equiv (1/\epsilon \beta )\ln (b/y_{0})\sim \beta ^{-1}$.
If, on the other hand, $\nu >1$ and the orbit is far enough from
the origin that $y^{\nu }\gg \beta y$, then $\beta $ can be neglected
and the time interval equals $T_{2}\equiv \left(y_{0}^{1-\nu }-b^{1-\nu }\right)/\left[\epsilon (\nu -1)\right]$.
For example, for $\epsilon =1.5,\beta =1,\nu =1.2,b=1,y_{0}=10^{-4}$
we find $T_{1}=6.1$ and $T_{2}=18$.

We have considered other models in which $g_{0}(y)$ is linear in
$y$ near $y=0$ but behaves as $\epsilon y^{\nu }$ for large $y$.
The cases investigated were $g_{0}(y)=\epsilon y(\beta ^{p}+y^{p(\nu -1)})^{1/p}$
for various values of $p$, including $p=1$. {[}Note that $g_{0}$
is analytic at $y=0$ if $p(\nu -1)$ is an integer.{]} The results
for all the tested values of $p$ are similar to the $p=1$ case described
above.

We have also considered the case $g_{0}(y)=\epsilon (\beta y+y^{2})$.
In Sec.~2 we concluded that the deterministic system for $\nu =2$
continued to have bursts of increasing amplitude and time interval
(rather than being capable of going to infinity in finite time in
a single burst) if $\epsilon <2$. Results for various values of $\epsilon <2$
and $D$ show that the results are similar to those for $\nu <2$,
as long as $\epsilon $ is small enough, $\beta $ is large enough,
and $D$ is large enough. Note that for this case the flow is analytic
everywhere, including $y<0$, and that for the deterministic form
there is a fixed point at $x=0,\: y=-\beta $, as well as the fixed
point at the origin, the latter having the $x-$axis as its stable
manifold. This new fixed point is attracting in both directions, and
therefore any noise in the $y-$direction eventually leads the orbit
to this fixed point.

\subsection{Limitation for large $y$ with asymmetry in $x$}

A phase portrait for a flow including both saturation in $y$ and
symmetry breaking in $x$ {[}c.f. eqs.~(\ref{eq:with-offset}), (\ref{eq:odell-2}){]},\[
\frac{dx}{dt}=(y-1)x+a,\]
\[
\frac{dy}{dt}=\epsilon y^{\nu }(1-\eta y)-x^{2}y,\]
is shown in Figure 14, with $\eta =0.1$ and $a=0.015$. %
\begin{figure}
\begin{center}\includegraphics[  width=4.5in,
  keepaspectratio]{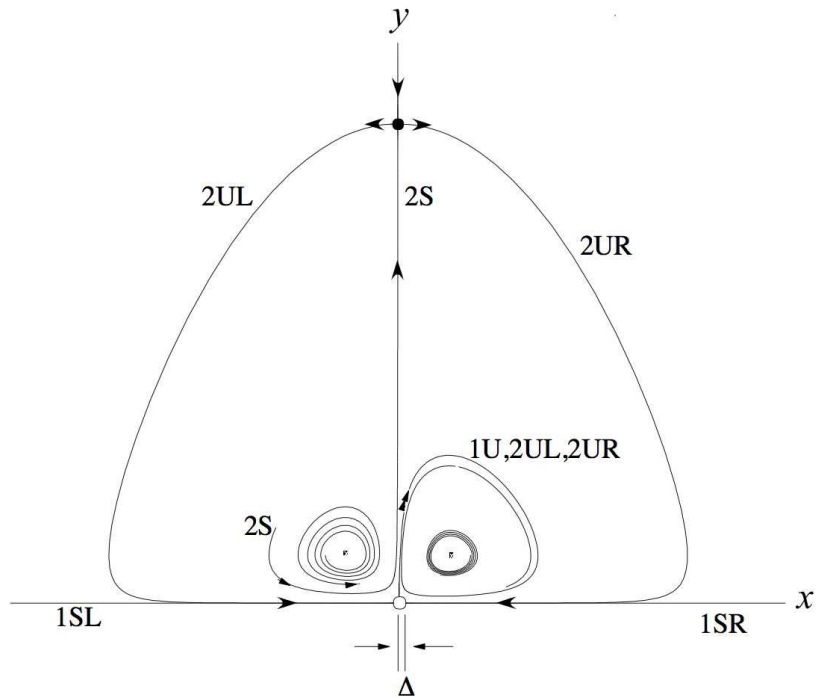}\end{center}

\caption{Phase portrait for the deterministic system of equations with limiting
in $y$ and offset in $x$, with zero noise and $\nu =1.2,\: \epsilon =0.5,\: a=0.015,\: \eta =0.1$.
There is a saddle near the origin (open circle), a saddle near $y=1/\eta $
(filled circle), \textcolor{black}{two unstable spirals, a}nd a stable
limit cycle (not shown) on the right. The symbols $1SL,\: 1SR$, and
$1U$ represent the left and right arms of the stable manifold of
the fixed point near the origin and its unstable manifold. The lower
arm of the stable manifold of the fixed point near $y=1/\eta $ is
$2S$ and its unstable manifold is $2UL,\: 2UR$. Points from the
\textcolor{black}{spiral} on the right go to the stable limit cycle;
points from the fixed point on the left eventually end up outside
$2S$ and go to the same limit cycle.}
\end{figure}
The unstable spirals are now slightly asymmetric due to the finite
value of $a$. There are two saddle points near the origin, at $x=a,\: y=0$
and at $x\approx a,\: y\approx (a^{2}/\epsilon )^{1/(\nu -1)}$. For
these parameters, the second fixed point has $y\sim a^{10}/\epsilon ^{5}\sim 10^{-17}$,
and the dynamics of the system can be described as if only the saddle
at $x=a,\: y=0$ exists. This saddle still has the $x-$axis as its
stable manifold (with right and left pieces labeled $1SR$ and $1SL$).
The unstable manifold of this saddle (labeled $1U$) now is no longer
the $y-$ axis, but bends slightly to the right and eventually asymptotes
to a limit cycle (not shown) orbiting the right spiral fixed point.
Another saddle at approximately $x\sim -a\eta ,\: y\sim 1/\eta $
(filled circle) has an unstable manifold with right and left pieces
(labeled $2UR$ and $2UL$, respectively). The invariant manifolds
bend downward, coming into the vicinity of the $x-$axis, pass very
close to the saddle at the origin, and both converge onto the unstable
manifold $1U$, thus approaching the limit cycle on the right as well.
The stable manifold for the upper saddle point, labeled $2S$, if
followed backward in time, asymptotes to the spiral on the left. Hence,
a narrow region on the $y-$axis near $y=1$ that is bounded by $2S$
on the left and $1U$ on the right sets the scale for the noise response.
If the noise amplitude is smaller than the width of this region (denoted
$\Delta $), nearly all points passing through this region will go
to the right and asymptote to the limit cycle. If $\sigma _{x}>\Delta $,
then orbits will get kicked to the left and right with nearly equal
probability, leading to noise stabilized behavior that prevents the
relaxation onto the limit cycle.

Thus, the presence of the symmetry breaking term in the deterministic
$dx/dt$ equation destroys the heteroclinic connection between the
two saddle points, leading generically to a limit cycle either on
the right or left, depending upon the sign of the offset. Thus, the
deterministic dynamics for $a=0,\eta >0$ discussed in Sec.~4.2 is
not structurally stable, but the behavior with noise is structurally
stable in the sense discussed at the end of Sec.~4.1. The noise response
is very similar to the noise response of the model with $\eta =a=0$
(Sec.~3), to the model with $\eta =0,a\neq 0$ (Sec. 4.1) and to
the model with $\eta >0,a=0$ (Sec.~4.2).

\subsection{Sinusoidal perturbation}

We have integrated eqs.(\ref{eq:x-eq.}), (\ref{eq:y-eq.}) with a
sinusoidal term $\xi (t)=b\sin (\omega t)$ added to the $x-$equation
rather than random noise. We chose $\omega $ to be large enough so
that the sine goes through many cycles when the orbit is along the
$x-$axis, but large enough to avoid aliasing, i.e.~$\omega h<\pi $,
where $h$ is the time step. The sinusoidal and random forms of $\xi (t)$
are extremes of temporal driving, with quasiperiodic time dependence
and colored random time dependence as intermediate cases. In all such
cases the analysis of Sec.~3.2 indicates that the typical value of
$x$ at $y=y_{2}$ is the important factor. (See Sec.~3.2 and Fig.~10.)
This suggests that the Lyapunov exponent $h_{1}$ has validity in
all these cases. To explore this further, we have obtained results
for $\nu =1.2,\: \epsilon =0.5$, as in Fig.~6, and with various
values of $\omega $ and $b$. The results were found to be qualitatively
similar to those with noise, with a simple relation between $b$ and
$D$, showing that indeed the accumulated effect on $x$ at the time
$y=y_{2}$ is the determining factor. That is, $\sigma _{x}\sim b/\omega $
or $b/\omega \sim D^{1/2}/\epsilon ^{1/4}$. In particular, the behavior
of $<|x_{n}|>$, $<T_{n}>$ and $h_{1}$ are similar. Thus, the similarity
of the results with this deterministic non-autonomous system and the
nonlinear stochastic system (\ref{eq:x-with-noise}), (\ref{eq:y-with-noise})
lend credence to the idea that $h_{1}$ as defined in Sec.~2 and
used in Sec.~3.1 is the appropriate form of the Lyapunov exponent
for the stochastic system. It is known that a system with periodic
driving can be distinguished from an autonomous system or one with
more complex temporal driving by means of nonlinear symbolic time
series analysis\cite{NLSTSA}. This distinction is possible because
of definite dips in the conditional entropy of symbolic time series
when the sampling time equals the period $2\pi /\omega $\cite{NLSTSA}.
This condition distinguishes periodic driving from all other temporal
driving (autonomous, quasi-periodic, colored noise, white noise),
but does not distinguish the other possible varieties from each other.
This topic is outside the scope of the present investigation.

\section{Electronic circuit}

In order to test for noise stabilization in a physical system, we
have constructed a circuit which integrates eqs.~(13) and (14). \textcolor{black}{I}n
dimensionless integral form, these equations are $x(\tau )=x_{0}+\int _{\tau _{0}}^{\tau }{\left(({y-1})x+\hat{\xi }(\tau ')\right)}d\tau '$
and $y(\tau )=y_{0}+\int _{\tau _{0}}^{\tau }{\left(\epsilon y^{\nu }-x^{2}y\right)}d\tau '$,
and the parameter \textcolor{black}{values} used in the circuit were
$\epsilon =0.5$ and $\nu =1.2$, as in Figs.~3,5-9,11,12. The circuit
design is shown in Fig.~18. The white noise, $\hat{\xi }(t)=\sqrt{2D}\xi (t)$,
stabilized the oscillations, and Figs.~15-17 show that the circuit
output agreed well with numerical solution of eqs.~(13) and (14).
We also observed the structural instability in these equations. See
Appendix B for a description of the circuit design. 

\begin{figure}
\begin{center}\includegraphics[  width=3in,
  keepaspectratio]{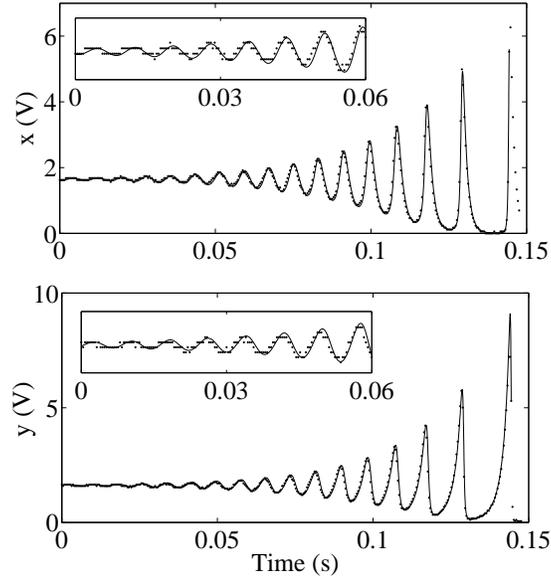}\end{center}

\caption{Circuit output (dots) compared to numerical solution of the ODE (traces),
with parameters as in Fig.~3. Adjusting the simulation parameters
to fit the data showed that all circuit parameters are within 3\%
of their expected values. The insets show the agreement of the \textcolor{black}{(digitized)}
data and simulation near the fixed point.}
\end{figure}

\subsection{Properties of the added noise}

The noise was generated by creating random numbers and recording them
to a \texttt{.wav} file to play back via the computer's audio output
at the standard rate of 44 kHz. This net process effectively filters
the noise through a lowpass filter. When we sampled the noise using
a digital oscilloscope, we found that the noise had a relatively constant
spectrum to frequencies as high as 20 kHz. We autocorrelated the noise,
and found that it was well represented by: \[
\left\langle V_{N}(t)V_{N}(t')\right\rangle =\frac{A_{0}}{\pi (t-t')}\sin {2\pi \frac{(t-t')}{T}}\]
 with a period $T=50$ $\mu $s, which also represents a flat spectrum
filtered by a 20 kHz low-pass filter. For times longer than $T/(2\pi )$,
this autocorrelation function is a good approximation of $A_{0}\delta (t)$.
By evaluating the autocorrelation function at $t=0$, we can determine
that $A_{0}=\frac{T}{2}\left\langle V_{N}^{2}\right\rangle $ so the
diffusion rate is $\frac{A_{0}}{2}$ or \[
D=\left\langle \left(\frac{V_{N}}{V_{2}}\right)^{2}\right\rangle \left(\frac{R_{2}}{R_{4}}\right)^{2}\frac{T}{4R_{1}C_{1}}\]
 in terms of the scaled variables used in Appendix B. The theoretical
minimum diffusion constant for our circuit parameters given by eq.~(\ref{eq:dmin_est})
is well below the intrinsic noise in the circuit. This intrinsic noise
is not well characterized and occurs in both the $x$ and $y$ variables.
We use a large enough value of the noise amplitude so that the intrinsic
noise contribution is negligible. We show in Figs.~16 and 17 the
quantities $T_{n-1}$ vs $x_{n}$ and $T_{n}$ vs $x_{n}$, first
obtained from the experiment and also by integrating numerically the
differential equations with the same parameters, in particular $D=4.7\times 10^{-4}$.
(These results are similar to those in Fig.~8, but with a different
value of $D$.) The agreement is very good.%
\begin{figure}
\begin{center}\includegraphics[  width=3in,
  keepaspectratio]{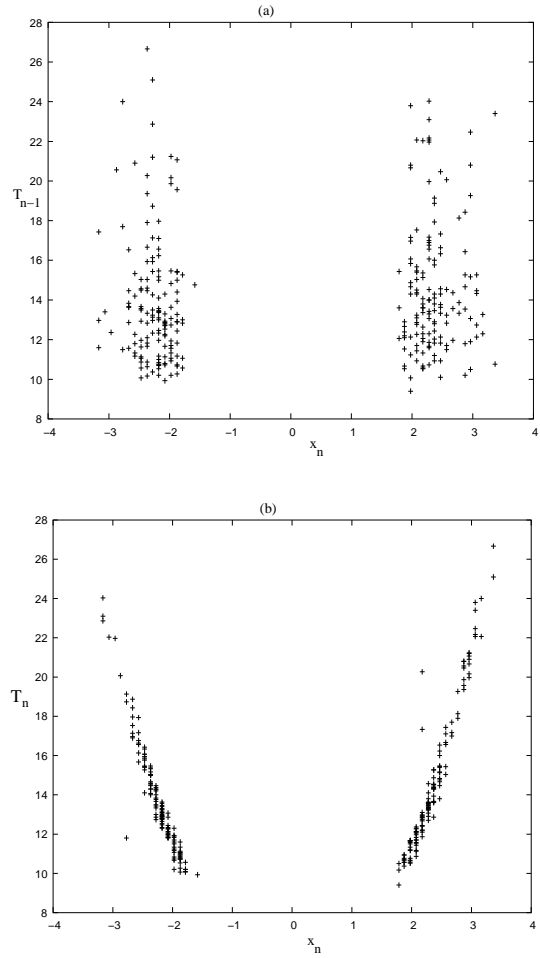}\end{center}

\caption{Comparison of peak height $x_{n}$ to (a) time since previous peak
$T_{n-1}$ and (b) time until next peak $T_{n}$, from experiments.
The correlations seen here are indicative of noise stabilization.
The noise level is $D\simeq 4.7\times 10^{-4}$. }
\end{figure}
\begin{figure}
\begin{center}\includegraphics[  width=3in,
  keepaspectratio]{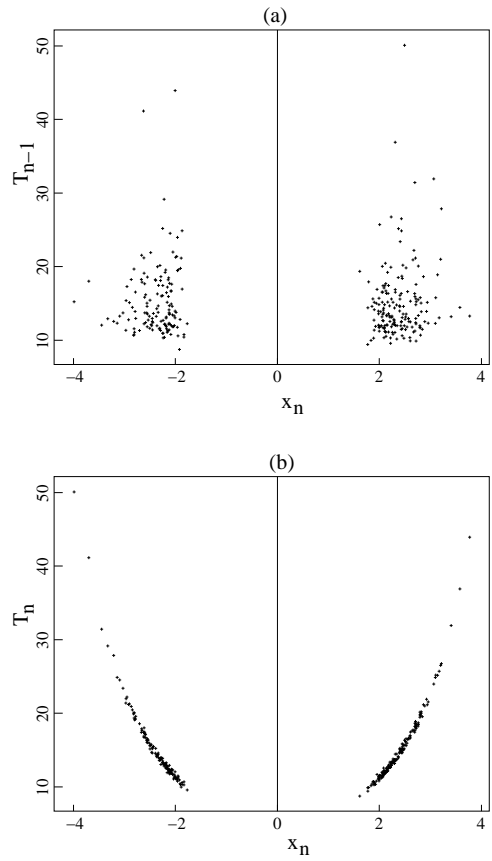}\end{center}

\caption{The same quantities as in Fig.~16 from numerical computation of
eqs.~(\ref{eq:x-with-noise}), (\ref{eq:y-with-noise}).}
\end{figure}

\subsection{Offsets and symmetry breaking}

The primary difficulty in designing this circuit is that small DC
offsets at the input of the integrators significantly change the differential
equations. In particular, an offset in the input to the $y$-integrator
either drives the $V_{y}$-output negative to create an error in the
AD538 computational unit, or it leads to a stable limit cycle similar
to that described in Sec.~4.4. We adjusted a small current ($\sim 0.45\mu $A)
to minimize the $V_{y}$-offset, using the automatic reset circuit
to recover whenever $V_{y}$ became negative. The reset kicks the
circuit back into the vicinity of one of the \textcolor{black}{unstable
spirals}. The $x$-integrator naturally follows, bringing $V_{x}$
to a value near its fixed point. Without this reset, a negative value
of $V_{y}$ leading to the failure of the AD538 causes the circuit
to fall to a stable fixed point with a large negative value of $V_{y}$.
An external trigger can also reset the circuit to values near its
unstable fixed point.

Similarly, we also corrected the offset in the $x$-integrator by
adding $\sim 0.2$ $\mu $A at the integrator input. We adjusted this
value until the noise signal generated equal numbers of negative and
positive $x$ pulses. After these adjustments, we observed the basic
structure of the oscillations as they evolved away from the fixed
point, in order to verify that the circuit waveforms were the same
as the model calculations (see Fig.~15). The fact that such a simple
adjustment can give results in agreement with the symmetric model
is consistent with the extended concept of structural stability discussed
at the end of Sec.~4.1. The results also show that the circuit is
a sensitive detector of offsets.

\section{Summary}

We have performed a study of a nonlinear stochastic ODE whose deterministic
form has unstable spirals, leading to bursty behavior, with successive
bursts growing in magnitude and with larger time intervals between
them. This bursty behavior is due to the fact that after each burst,
the orbit comes closer to the unstable manifold ($y-$axis) of a hyperbolic
fixed point at the origin, and therefore travels farther along this
unstable manifold before diverging from it to form the next burst.

In the presence of noise at a very small level, the bursts get stabilized
in the sense of becoming limited in magnitude\textcolor{black}{.}
The time interval between them also limited\textcolor{black}{,} and
the bursts can go to either positive or negative $x$\textcolor{black}{.}
In many qualitative senses, the behavior appears like deterministic
chaos.

This system has reflection symmetry in $x$; an offset $a$ in $x$
destroying this symmetry can lead to completely different behavior,
depending on its magnitude relative to the noise. That is, the bursty
behavior seen in the symmetric deterministic equations is not structurally
stable. With noise and a small value of the offset $|a|<\sqrt{2D}$
($D$ is the Brownian diffusion coefficient), the bounded bursty behavior
persists, but with more bursts going to the right if $a>0$ (to the
left if $a<0$.) For larger offset $a\gtrsim \sqrt{2D}$, all bursts
go to the right and basically give a noisy form of the stable limit
cycle. In this sense, the results in the presence of noise and $a=0$
are structurally stable.

We have considered modifications to the model allowing for saturation
of $y$, because bursts cannot continue to grow without bound in a
physical system. We have also considered modifications near the saddle
at the origin, to give the saddle at the origin a positive eigenvalue.
This change in the linear part of the flow near the saddle affects
the time intervals between bursts, making their characteristic value
much smaller, but does not affect the properties of the burst amplitudes,
or the signs (in $x$) of the bursts.

We have described briefly results on a nonlinear circuit satisfying
the same equations as the model. The circuit behaves similarly to
the model. In particular, the circuit is very sensitive to the presence
of an offset, and in practice the offset is adjusted to minimize the
asymmetry of the signal. More details are presented in Ref.~\cite{Conference-on-Expermental-Chaos}
and in Appendix B.

The system (\ref{eq:x-with-noise}), (\ref{eq:y-with-noise}) and
its generalizations in Sec.~4 are arguably the simplest realizations
of systems in which a small noise level can limit the amplitude of
bursts and lead to qualitatively distinct behavior. We have listed
in the Introduction physical examples of systems in which this effect
may be important. For the tokamak example, the results here should
have an impact on low dimensional modeling of ELMs. Indeed, the observation
of chaotic time dependence of ELM data suggests that a simple autonomous
ODE model must be three-dimensional. However, tokamaks are known to
have a broad spectrum of fluctuations (turbulence). If these fluctuations
can be treated as uncorrelated noise, i.e.~if their correlation time
is much shorter that ELM time scales, it is justifiable to explore
two-dimensional models with noise such as the models studied here.

\section*{\textcolor{black}{Appendix A: Fokker-Planck Equation}}

\textcolor{black}{\setcounter{equation}{0}}

\renewcommand{\theequation}{A-\arabic{equation}}

The stochastic behavior of eq.~(\ref{eq:linear-stochastic}) is governed
by the Fokker-Planck equation for the probability density function
$f(x,t)$,\begin{equation}
\frac{\partial f}{\partial t}+\frac{\partial }{\partial x}\left(\gamma (t)xf\right)=\frac{\partial }{\partial x}\left(D\frac{\partial f}{\partial x}\right),\label{eq:fokker-planck}\end{equation}
where $D=\sigma ^{2}/2$ is the diffusion coefficient. For arbitrary
$\gamma (t)$, (\ref{eq:fokker-planck}) has the exact solution\[
f(x,t)=\sqrt{\frac{1}{2\pi \alpha (t)}}e^{-x^{2}/2\alpha (t)}\]
if the variance or temperature $\alpha (t)$ satisfies\textcolor{black}{\begin{equation}
\frac{d}{dt}\alpha (t)=2\left[\alpha (t)\gamma (t)+D\right].\label{eq:alpha-equation}\end{equation}
Eq}.~(\ref{eq:alpha-equation}) has the solution \[
\alpha (t)=2D\int _{-\infty }^{t}ds_{1}e^{2\int _{s_{1}}^{t}\gamma (s_{2})ds_{2}},\]
 assuming $\alpha (t\rightarrow -\infty )=0$. Thus, $\alpha (t)$
is proportional to $D$, with a coefficient depending on $\gamma (t)$. 

If $\gamma $ is approximately constant ($|\dot{\gamma }/\gamma ^{2}|\ll 1$)
and negative, $\alpha $ approaches a slowly varying state with $\alpha (t)=D/|\gamma (t)|$,
in which the inward motion due to the advective term in (\ref{eq:fokker-planck})
balances diffusion and $\partial f/\partial t$ is negligible. This
limit gives\begin{equation}
f(x,t)\rightarrow \sqrt{|\gamma |/2\pi D}e^{-|\gamma |x^{2}/2D}.\label{eq:FP-constant-gamma}\end{equation}
Another limit is recovered by neglecting $\gamma (t)$ in eq.~(\ref{eq:fokker-planck}),
giving \[
\alpha (t)=\alpha (t_{1})+2D(t-t_{1})=2Dt+2\alpha _{0},\]
\textcolor{black}{where $t_{1}$ is the time when $\dot{\gamma }\sim \gamma ^{2}$.}
Without loss of generality we can set the time where $\gamma =0$
to $t=0$. This range, in which the advective term in eq.~(\ref{eq:fokker-planck})
is small, gives the purely diffusive random walk result \begin{equation}
f(x,t)\sim \frac{1}{\sqrt{4\pi (Dt+\alpha _{0})}}e^{-x^{2}/4(Dt+\alpha _{0})}.\label{eq:FP-gamma=0}\end{equation}
A third range has $\gamma $ positive with advection dominating diffusion.
We find \begin{equation}
\alpha (t)=\alpha (t_{2})\exp \left(2\int _{t_{2}}^{t}\gamma (s)ds\right),\label{eq:alpha-late}\end{equation}
where $t_{2}$ is the time this range is entered, i.e.~where $\gamma (t_{2})\alpha (t_{2})\sim D$.
In this range the noise becomes negligible. 

\textcolor{black}{A simple example having these properties has $\gamma $
linear in time, $\gamma (t)=\dot{\gamma }_{0}t$.} Again taking $\alpha (t=-\infty )=0$,
we find \[
\alpha (t)=2De^{\dot{\gamma }_{0}t^{2}}\int _{-\infty }^{t}e^{-\dot{\gamma }_{0}s^{2}}ds.\]
In this example $\alpha (t)$ has slow growth for $t<t_{1}\equiv -1/\sqrt{\dot{\gamma }_{0}}$,
diffusive increase for $t_{1}<t<t_{2}$, where $t_{2}=1/\sqrt{\dot{\gamma }_{0}}$,
and exponential growth for $t>t_{2}$. The value of $\alpha (t)$
at $t=0$ (corresponding to $y=1$) is $\sigma _{x}^{2}\equiv \alpha (0)\sim D/\sqrt{\dot{\gamma }_{0}}$.

For application to eqs.~(\ref{eq:x-with-noise}), (\ref{eq:y-with-noise}),
consider $x$ small so that its equation is linear (when the second
term on the right in (\ref{eq:y-with-noise}) is negligible). We then
note that if $\alpha $ is small for $y\approx 0$, then $\alpha (t)$
\textcolor{black}{near $y=1$ (recall $\gamma (t)=y(t)-1=0$) is}
proportional to $D/\sqrt{\dot{\gamma }_{0}}$. Since $\dot{\gamma }=\dot{y}\sim \epsilon $,
we have $\alpha (y\approx 1)\sim D/\sqrt{\epsilon }$. After a diffusive
stage, $\alpha $ continues to increase as in eq.~(\ref{eq:alpha-late}),
with noise no longer playing a role. Thus, the nonlinear orbit for
later times depends only on the noise accumulated by the time (here
$t=t_{2}$) just after the orbits cross the throat at $y=1$; the
value of $x$ at $y\approx y_{2}$, when noise last plays a role,
is proportional to $\sqrt{\alpha }\propto D^{1/2}/\epsilon ^{1/4}$.
See Fig.~10. Thus, in essence, the orbit from the crossing of $y=1$
with small $x$ out to the next crossing and back to near the origin
is deterministic, and the noise plays its role only along the $y-$axis.

\section*{Appendix B: Circuit Design}

\textcolor{black}{\setcounter{equation}{0}}

\renewcommand{\theequation}{B-\arabic{equation}}

The design of our circuit is basically the same as reported in Ref.~\cite{Conference-on-Expermental-Chaos},
but we have adjusted our circuit parameters, and extended the analysis
of the circuit behavior. For the sake of completeness, we have included
all of the new circuit parameters in this appendix, as well as our
analysis of the minimum noise amplitude necessary to keep the circuit
from saturating the circuit elements.

The analog circuit consists of three basic sub-circuits: the $x$-integrator,
the $y$-integrator, and the reset controller, as shown in Fig.~18.
\begin{figure}
\begin{center}\includegraphics[  width=6in,
  keepaspectratio]{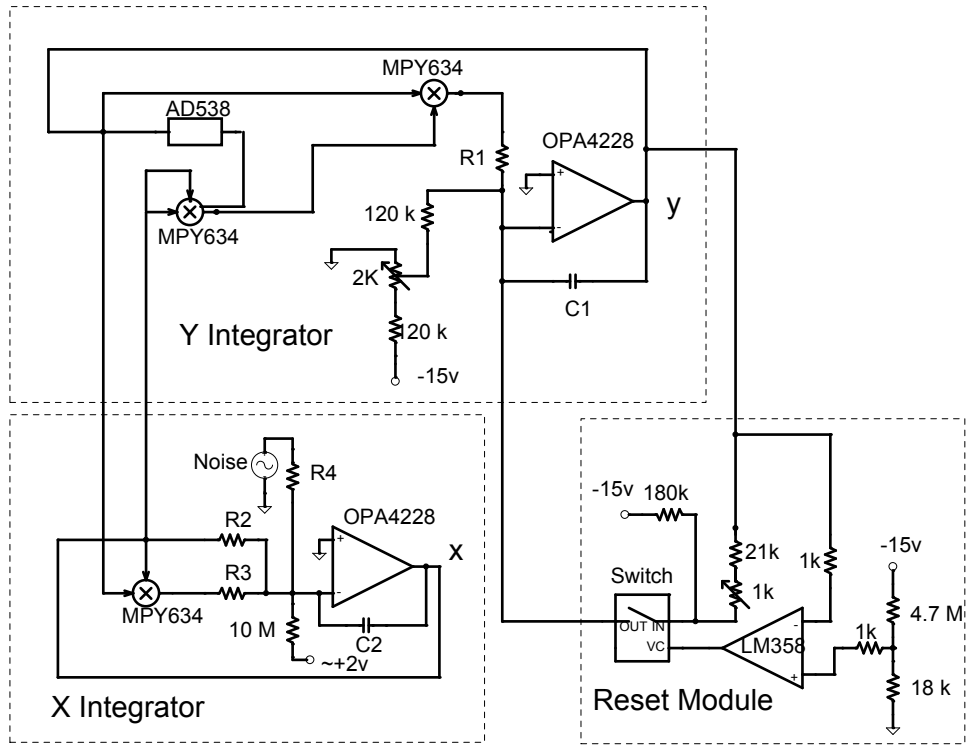}\end{center}

\caption{Circuit diagram.}
\end{figure}
The integrators use OPA 4228 operational amplifiers (low noise, 33
MHz bandwidth) with capacitive feedback (10 nF) to integrate their
inputs. $V_{1}$ and $V_{2}$ are constant applied voltages, while
$V_{x}$ and $V_{y}$ are time varying voltages, proportional to $x(\tau )$
and $y(\tau )$, respectively. 

The input to the $y$-integrator uses an AD538 real-time computational
unit (400 kHz bandwidth) to raise the $V_{y}$ voltage to a fractional
power, $V_{y}(t)^{\nu -1}$, by taking its logarithm, scaling the
result by $\nu -1$, and then exponentiating to generate $V_{1}(V_{y}(t)/V_{2})^{\nu -1}$.
This output is then added into the output of an MPY634 precision multiplier
(10 MHz bandwidth) that creates the ratio $V_{x}^{2}(t)/V_{2}$. A
second MPY634 multiplies this combined signal by $V_{y}/V_{2}$ before
it enters the integrator. We also use additional small adjustable
current sources to eliminate offsets. 

The input to the $x$-integrator is the sum of $V_{x}$, the noise
source, and $V_{x}V_{y}/V_{2}$ formed by another MPY634. The net
output signal of the entire circuit has a maximum frequency of 2 KHz,
well within the bandwidth limit of all the components. This circuit
does the following integrations:\[
\begin{array}{l}
 V_{x}(t)=V_{x}(t_{0})+\int _{t_{0}}^{t}{\left(\frac{R_{2}V_{y}(t')}{R_{3}V_{2}}-1\right)V_{x}(t')\frac{dt'}{R_{2}C_{2}}}+\int _{t_{0}}^{t}{V_{N}(t')\frac{dt'}{R_{4}C_{2}}},\\
 V_{y}(t)=V_{y}(t_{0})+\int _{t_{0}}^{t}{\left(V_{1}\left(\frac{V_{y}(t')}{V_{2}}\right)^{\nu }-{\left(\frac{V_{x}(t')}{V_{2}}\right)}^{2}V_{y}(t')\right)\frac{dt'}{R_{1}C_{1}}},\end{array}
\]
where the circuit components had the values listed in Table 1, and
the parameter $\nu -1$ was set to 0.2 in the AD538 component by a
voltage divider composed of a 2200 $\Omega $ resistor and a 560 $\Omega $
resistor. This dimensional form of the equations is related to the
dimensionless form by defining $x$, $y$, $\tau $, $\epsilon $
and $\eta $ as:

\[
\begin{array}{l}
 y=\frac{R_{2}}{R_{3}}\frac{V_{y}}{V_{2}}\\
 \tau =\frac{t}{R_{2}C_{2}}\\
 \epsilon =\frac{R_{2}C_{2}}{R_{1}C_{1}}\frac{V_{1}}{V_{2}}\left(\frac{R_{3}}{R_{2}}\right)^{\nu -1}\\
 x=\sqrt{\frac{R_{2}C_{2}}{R_{1}C_{1}}}\frac{V_{x}}{V_{2}}=\sqrt{\epsilon }\frac{V_{x}}{\sqrt{V_{1}V_{2}}}\left(\frac{R_{2}}{R_{3}}\right)^{\frac{\nu -1}{2}}\\
 \eta =\sqrt{\frac{R_{2}C_{2}}{R_{1}C_{1}}}\frac{V_{N}}{V_{2}}\frac{R_{2}}{R_{4}}\end{array}
\]

This leads to fixed points at: \[
\begin{array}{l}
 V_{y*}=\frac{R_{3}}{R_{2}}V_{2}\\
 V_{x*}=\sqrt{V_{1}V_{2}}{\left(\frac{R_{3}}{R_{2}}\right)}^{\frac{\nu -1}{2}}\end{array}
\]
 Thus, a circuit design with a given value of $\epsilon $ has its
fixed points and its voltage scaling determined by the choice of the
ratio $R_{3}/R_{2}$. This value can be optimally set by forcing both
the $x$ circuit and the $y$ circuit to reach saturation values on
the same cycle. For the $\nu =1$ case, neglecting the logarithmic
terms of the Hamiltonian $H(x,y)$ in eq.~(\ref{eq:hamiltonian}),
the peak value of $y$ ($y_{p}$) and its following peak value of
$x$ ($x_{p}$) are related by $x_{p}^{2}=2y_{p}$ if $H$ is large
enough, i.e.~for bursts with $x_{p},\: y_{p}$ large enough. These
two peak values cannot require voltages in excess of $V_{2}$, or
the multipliers will \textcolor{black}{fail,} \textbf{\textcolor{black}{}}\textcolor{black}{and
the peaks will be clipped.} To optimize, we equate these peaks when
they reach $V_{2}$; for the $\nu =1$ case this gives $\epsilon V_{2}/V_{1}=2R_{2}/R_{3}$
or, for our values of $V_{1}$ and $V_{2}$,\[
\frac{R_{2}}{R_{3}}=\frac{\epsilon }{2}\frac{V_{2}}{V_{1}}=6.25.\]

This choice then implies maximum values of $x_{m}=\sqrt{2\left(\epsilon V_{2}/2V_{1}\right)}=3.53$,
and $y_{m}=\epsilon V_{2}/2V_{1}=6.25$. These maximum values of $x$
and $y$ determine the minimum noise amplitude that must be present
to keep the voltage peaks within the operating range of the multipliers.
The logarithmic dependence observed in Fig.~11 can be approximated
as $\left\langle x\right\rangle =(1/8)\ln \left(10^{5}/D\right)$,
so that: \begin{equation}
D_{min}=10^{5}e^{-8x_{m}}=10^{5}e^{-8\sqrt{2\left(\frac{\epsilon V_{2}}{2V_{1}}\right)}}\sim 2\times 10^{-10}.\label{eq:dmin_est}\end{equation}
When the amplitudes are low enough to avoid clipping, the measured
results are in agreement with those given in Sec.~3.2.

\begin{table}
\[
\begin{array}{cc}
 V_{1} & 0.4V\\
 V_{2} & 10V\\
 R_{1} & {6.8k}\Omega \\
 R_{2} & {122k}\Omega \\
 R_{3} & {19.5k}\Omega \\
 R_{4} & {67k}\Omega \\
 C_{1} & 10nF\\
 C_{2} & 10nF\end{array}
\]

\caption{Values of circuit elements.}
\end{table}

\section*{ACKNOWLEDGMENTS}

We wish to thank C. Doering, J. Guckenheimer, E. Ott, and D. Sigeti
for valuable discussions. This work was supported by the U. S. Department
of Energy, under contract W-7405-ENG-36, Office of Science, Office
of Fusion Energy Sciences, and by the NSF-DOE Program in Basic Plasma
Physics under contract PHY-0317256.

\newpage
\bibliographystyle{prsty}
\bibliography{paper}
\newpage

\end{document}